\documentclass[twocolumn]{aastex62}
\usepackage{lineno}
\usepackage{bm}
\usepackage{csvsimple}
\usepackage{savesym}
\savesymbol{tablenum}
\usepackage{siunitx}
\restoresymbol{SIX}{tablenum}
\usepackage{booktabs,array}

\newcount\rowc

\makeatletter
\def\ttabular{%
\hbox\bgroup
\let\\\cr
\def\rulea{\ifnum\rowc=\@ne \hrule height 0.5pt \fi}
\def\ruleb{
\ifnum\rowc=1\hrule height 0.5pt \else
\ifnum\rowc=6\hrule height \lightrulewidth 
   \else \hrule height \lightrulewidth\fi\fi}
\valign\bgroup
\global\rowc\@ne
\rulea
\vspace{2pt}
\hbox to 9.9em{\strut \hfill##\hfill}%
\ruleb
\vspace{3pt}
&&%
\global\advance\rowc\@ne
\hbox to 9.9em{\strut\hfill##\hfill}%
\vspace{3pt}
\cr}
\def\endttabular{%
\crcr\egroup\egroup}

\newcommand{\eqb}{\begin{eqnarray}}
\newcommand{\eqe}{\end{eqnarray}}
\newcommand{\beq}{\begin{equation}}
\newcommand{\eeq}{\end{equation}}

\shorttitle{Millilensing of PKS1413+135}
\shortauthors{Peirson et al.}

\begin{document}

\title{New Tests of Millilensing in the Blazar PKS 1413+135}

\correspondingauthor{A. L. Peirson}
\email{alpv95@stanford.edu}

\author{A. L. Peirson}
\affiliation{Kavli Institute for Particle Astrophysics and Cosmology, Stanford University, Stanford, CA 94305, USA}
 \author{I. Liodakis}
\affiliation{Finnish Center for Astronomy with ESO, University of Turku, Vesilinnantie 5, FI-20014, Finland}
 \author{A.C.S Readhead}
\affiliation{Owens Valley Radio Observatory, California Institute of Technology,  Pasadena, CA 91125, USA}
\author{M.L. Lister}
\affiliation{Department of Physics and Astronomy, Purdue University, 525 Northwestern Avenue, West Lafayette, IN 47907, USA}
\author{E. S. Perlman}
\affiliation{Dept. of Physics and Space Sciences, Florida Institute of Technology, 150 W. University Boulevard, Melbourne, FL, 32901, USA}
\author{M.F. Aller}
\affiliation{Department of Astronomy, University of Michigan, 311 West Hall, 1085 S. University Avenue, Ann Arbor, MI 48109, USA}
\author{R. D. Blandford}
\affiliation{Kavli Institute for Particle Astrophysics and Cosmology, Stanford University, Stanford, CA 94305, USA}
% \author{P.J. Elwood}
% \affiliation{Astrophysics Group, Cavendish Laboratory, 19 J.J. Thompson Avenue, Cambridge CB3 0HE, UK}
 \author{K.J.B. Grainge}
\affiliation{Jodrell Bank Centre for Astrophysics, University of Manchester, Oxford Road, Manchester M13 9PL, UK} 
\author{D.A. Green}
\affiliation{Astrophysics Group, Cavendish Laboratory, 19 J.J. Thompson Avenue, Cambridge CB3 0HE, UK}
\author{M. A. Gurwell}
\affiliation{Center for Astrophysics | Harvard \& Smithsonian, Cambridge, MA 02138, USA} 
\author{M.W. Hodges}
\affiliation{Owens Valley Radio Observatory, California Institute of Technology,  Pasadena, CA 91125, USA} 
\author{T. Hovatta}
\affiliation{Finnish Center for Astronomy with ESO, University of Turku, Vesilinnantie 5, FI-20014, Finland}
\affiliation{Aalto University Mets\"ahovi Radio Observatory,  Mets\"ahovintie 114, 02540 Kylm\"al\"a, Finland}
\author{S. Kiehlmann}
\affiliation{Institute of Astrophysics, Foundation for Research and Technology-Hellas, GR-71110 Heraklion, Greece}
\affiliation{Department of Physics, Univ. of Crete, GR-70013 Heraklion, Greece} 
\author{A. L\"ahteenm\"aki}
\affiliation{Aalto University Mets\"ahovi Radio Observatory,  Mets\"ahovintie 114, 02540 Kylm\"al\"a, Finland} 
\author{W. Max-Moerbeck} 
\affiliation{Departamento de Astronomia, Universidad de Chile, Camino El Observatorio 1515, Las Condes, Santiago, Chile}
\author{T. Mcaloone}
\affiliation{Jodrell Bank Centre for Astrophysics, University of Manchester, Oxford Road, Manchester M13 9PL, UK} 
\author{S. O'Neill}
\affiliation{Owens Valley Radio Observatory, California Institute of Technology,  Pasadena, CA 91125, USA}
\author{V. Pavlidou} 
\affiliation{Department of Physics and Institute of Theoretical and Computational Physics, University of Crete, 71003 Heraklion, Greece}
\author{T. J. Pearson}
\affiliation{Owens Valley Radio Observatory, California Institute of Technology,  Pasadena, CA 91125, USA} 
\author{V. Ravi}
\affiliation{Owens Valley Radio Observatory, California Institute of Technology,  Pasadena, CA 91125, USA} 
\author{R.A. Reeves}
\affiliation{CePIA, Astronomy Department, Universidad de Concepci\'on,  Casilla 160-C, Concepci\'on, Chile} 
%\author{J.L. Richards}
%\affiliation{Owens Valley Radio Observatory, California Institute of Technology,  Pasadena, CA 91125, USA}
%\author{R.D.E. Saunders}
%\affiliation{Astrophysics Group, Cavendish Laboratory, 19 J.J. Thompson Avenue, Cambridge CB3 0HE, UK}
\author{P.F. Scott}
\affiliation{Astrophysics Group, Cavendish Laboratory, 19 J.J. Thompson Avenue, Cambridge CB3 0HE, UK}
\author{G.B. Taylor}
\affiliation{Department of Physics and Astronomy, University of New Mexico, Albuquerque, NM 87131, USA}
\author{D.J. Titterington}
\affiliation{Astrophysics Group, Cavendish Laboratory, 19 J.J. Thompson Avenue, Cambridge CB3 0HE, UK}
\author{M. Tornikoski}
\affiliation{Aalto University Mets\"ahovi Radio Observatory,  Mets\"ahovintie 114, 02540 Kylm\"al\"a, Finland} 
\author{H.K. Vedantham}
\affiliation{ASTRON, Netherlands Institute for Radio Astronomy, P.O. Box 2, 7990 AA Dwingeloo, the Netherlands}
\author{P.N. Wilkinson}
\affiliation{Jodrell Bank Centre for Astrophysics, University of Manchester, Oxford Road, Manchester M13 9PL, UK} 
\author{D. T. Williams}
\affiliation{Jodrell Bank Centre for Astrophysics, University of Manchester, Oxford Road, Manchester M13 9PL, UK} 
\author{J. A. Zensus}
\affiliation{Max-Planck-Institut f\"ur Radioastronomie, Auf dem H\"ugel 69, D-53121 Bonn, Germany}

\begin{abstract}

% Gravitational lensing of planets, supernovae, galaxies and more, allows us to study both distant or faint objects that would otherwise remain undetected, as well as the objects responsible for the lensing. 
Symmetric Achromatic Variability (SAV) is a rare form of radio variability in blazars that has been attributed to gravitational millilensing by a $\sim 10^2 - 10^5 M_\odot$ mass condensate. Four SAVs have been identified between 1980 and 2020 in the long-term radio monitoring data of the blazar PKS 1413+135. We show that all four can be fitted with the same, unchanging, gravitational lens model. If SAV is due to gravitational milli-lensing, PKS 1413+135  provides a unique system for studying active galactic nuclei with unprecedented $\mu$as  resolution, as well as for studying the nature of the millilens itself.
{We discuss two possible candidates for the putative millilens: a giant molecular cloud hosted in the intervening edge-on spiral galaxy, and an undetected dwarf galaxy with a massive black hole.}
% , which could well be a dark matter subhalo. 
We find a significant dependence of SAV crossing time on frequency, which could indicate a fast shock moving in a slower underlying flow. We also find tentative evidence for a 989-day periodicity in the SAVs, which, if real, makes possible the prediction of future SAVs: the next three windows for possible SAVs begin in August 2022, May 2025, and February 2028.
%We have constrained millilens parameters
\end{abstract}

\section{Introduction}\label{sec:introduc}

\begin{figure*}
\resizebox{\hsize}{!}{\includegraphics[scale=1]{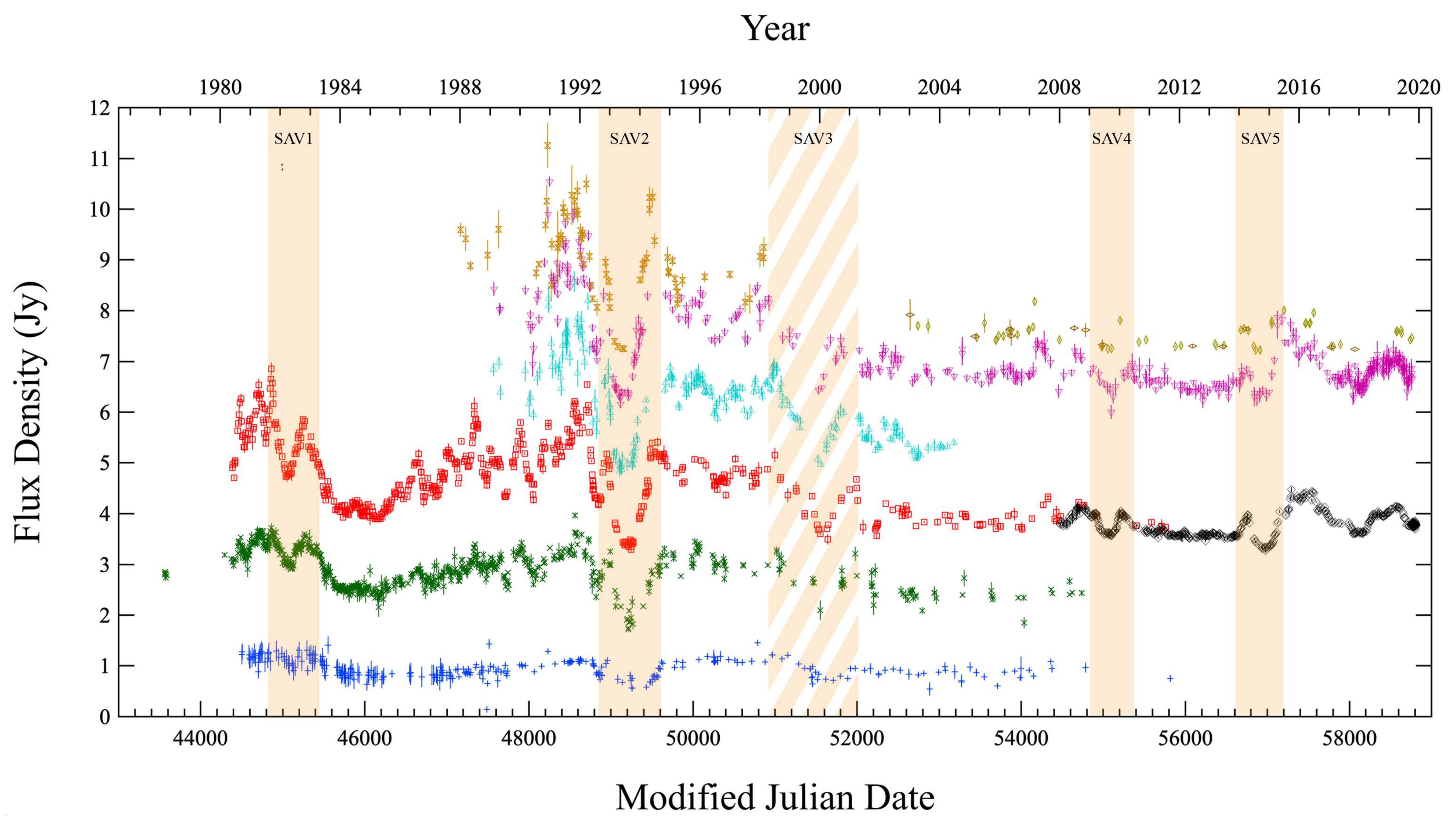} }
\caption{PKS~1413+135 radio light curves. The four bona fide SAVs discussed in this paper are indicated by the fully shaded areas. The cross-hatched area indicates the anomalous SAV3 event. The blue plusses show the 4.8 GHz UMRAO data. The other light curves have been shifted for clarity as  follows: UMRAO 8 GHz +1.5 Jy (green crosses), UMRAO 14.5 GHz +3 Jy (red squares), OVRO and AMI 15 GHz +3 Jy (black circles), MRO 22 GHz +4.5 Jy (cyan up triangles), MRO 37 GHz + 6 Jy (purple down triangles), IRAM 90 GHz + 7 Jy (brown xboxes), SMA 230 GHz + 7 Jy (khaki horizontal squashed diamonds), and SMA 345 GHz +7 Jy (khaki vertical squashed diamonds).}
\label{plt:lightcurves}
\end{figure*}

A new form of variability in blazars  ``Symmetric Achromatic Variability'' (SAV) was reported by \citet{Vedantham2017}, hereafter Paper 1,  and attributed to gravitational milli-lensing by a $\sim 10^2 - 10^5 M_\odot$ mass condensate.  In a  companion paper \citep{Vedantham2017-II}, hereafter Paper 2, the possibility that SAV might be due to Extreme Scattering Events (ESEs) \citep{1987Natur.326..675F,1994ApJ...430..581F} was definitively ruled out.  It has been shown that the blazar PKS 1413+135 is almost certainly located behind an edge-on spiral Seyfert 2 galaxy at z=0.247 and that the blazar lies in the redshift range $0.247<z<0.5$ \citep{2021ApJ...907...61R}, hereafter Paper 3. If PKS 1413+135 does indeed lie behind the Seyfert 2 galaxy, then the galaxy provides a natural host for the putative mass condensate responsible for the milli-lensing. However, if PKS 1413+135 is located in the spiral galaxy, then, as pointed out in Paper 1, any putative milli-lens would be an intergalactic mass condensate of mass $\sim 10^2 - 10^5 M_\odot$, with this population having $\Omega_l/\Omega_m = 10^{-1} - 10^{-3}$. Interestingly these ranges of mass and cosmological density overlap the mass and cosmological density deduced by \citet{2021NatAs.tmp...55P} for the intermediate mass condensate they have possibly detected through a lensed gamma-ray burst.  In total five candidate SAVs have been identified in the radio light curves of the active galactic nucleus (AGN) PKS 1413+135 between 1980 and 2020 (see Fig. \ref{plt:lightcurves}).

In Papers 1 \& 3 a model was suggested comprised of a stationary, unvarying lens and a background jetted-AGN that sends successive high-speed ($v \sim c$) components across the field behind the lens, resulting in repeated time-variable lensing features in the AGN light curves. {The lens is stationary in the sense that it moves very little on timescales of decades.  Since the lens is not in the Galaxy (Paper 1. Fig. 6) it would need to be traveling at relativistic speed in order to move significantly on timescales of decades.}  

If a blazar exhibits SAV, and if SAV is due to milli-lensing, the blazar can be monitored for subsequent SAVs, which can be studied in detail via multi-wavelength campaigns as well as by very long baseline interferometry (VLBI). In addition, repeated SAVs provide independent probes of the lensing system,  enabling us to refine the lens model, in contrast to the gravitational lensing of gamma-ray bursts, which  do not repeat.
% We note that if the stationary lens is located in the spiral galaxy
% Although we observe some chromaticity in the source components due to ,

The achromaticity and symmetry of SAV are easily explained by gravitational lensing, which will also amplify the variability and the fraction of the total flux density in the unresolved radio core.
%All of these properties of PKS~1413+135 have a natural explanation under the gravitational lensing hypothesis.

The full details of the lens required to explain the SAV in PKS~1413+135 are given in Paper 1, so we do not repeat them  here. As was shown, the lens cannot be located in our own galaxy because this implies surface densities of $\gtrsim 10^4 M_\odot \; {\rm pc}^{-2}$, and such concentrations are not found in our galaxy. However, assuming that PKS~1413+135 is at redshift $z \sim 0.5$, i.e. roughly  450 Mpc behind the lens, on the assumption that  the lens is associated with the spiral galaxy, then the projected density needed for strong lensing is $\sim 5 \times 10^3 M_\odot {\rm pc}^{-2}$ (Paper 1, Fig. 6).  

\begin{figure*}[ht!]
\centering
\includegraphics[width=1.0\textwidth]{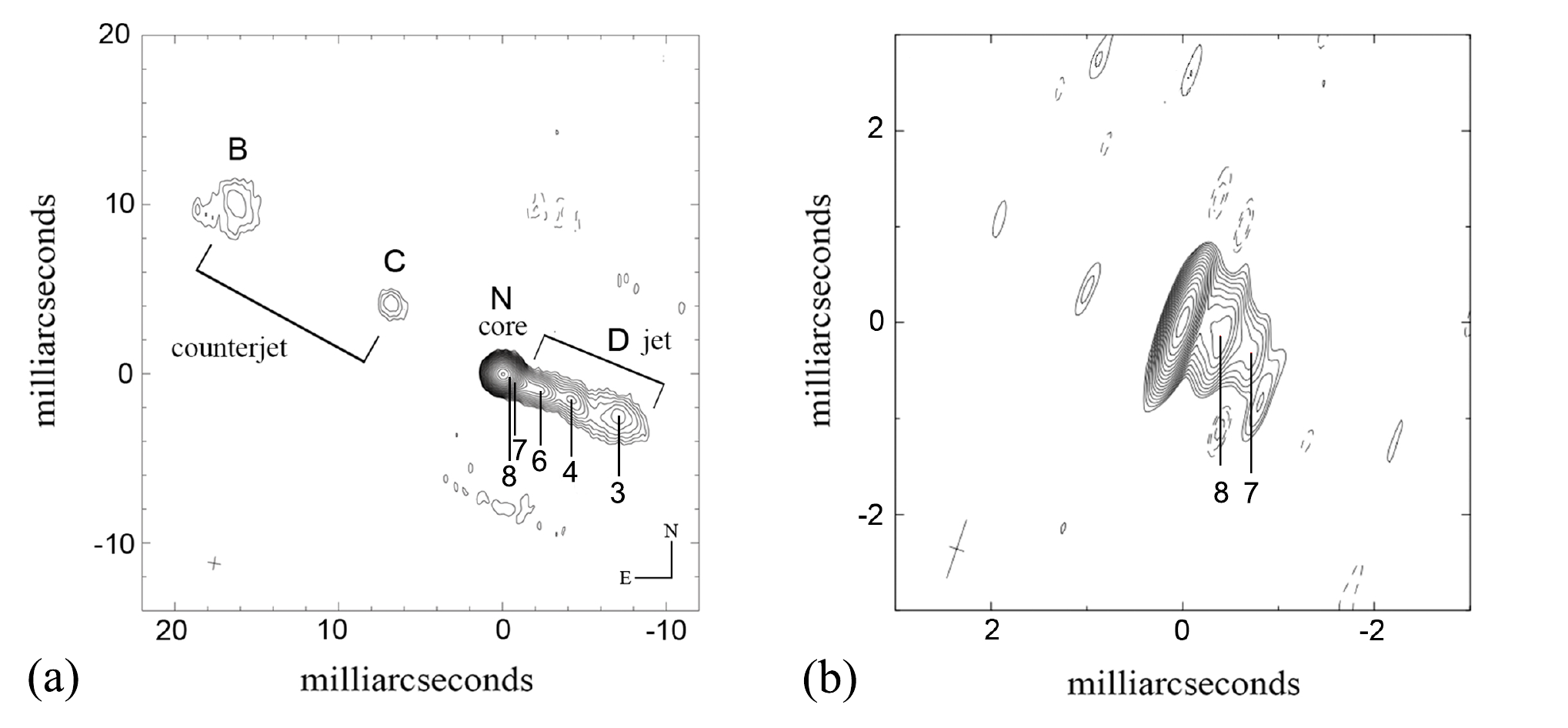}
\caption{The inner 30 milliarcseconds of the radio structure of PKS 1413+135. (a)  stacked MOJAVE 15 GHz image made from maps at 23 epochs showing the component numbering system adopted by \citet{2019ApJ...874...43L} and used in this paper. Nine of these epochs coincided with SAV3. (b) the 43 GHz map made by the MOJAVE team from archival observations (VLBA code BP048D; P.I. Eric Perlman) taken during SAV4, for which the restoring beam (FWHM) is $0.64 \times 0.16$ millarcseconds$^2$ in PA $-18.7^\circ$. No evidence has been found of multiple stationary images of the core that could be caused by gravitational millilensing on scales down to 0.25 mas, placing an upper limit of 0.25 milliarcseconds  on the Einstein radius of the putative millilens. }
\label{plt:structure}
\end{figure*}

\begin{figure}[ht!]
\centering
\includegraphics[width=0.5\textwidth]{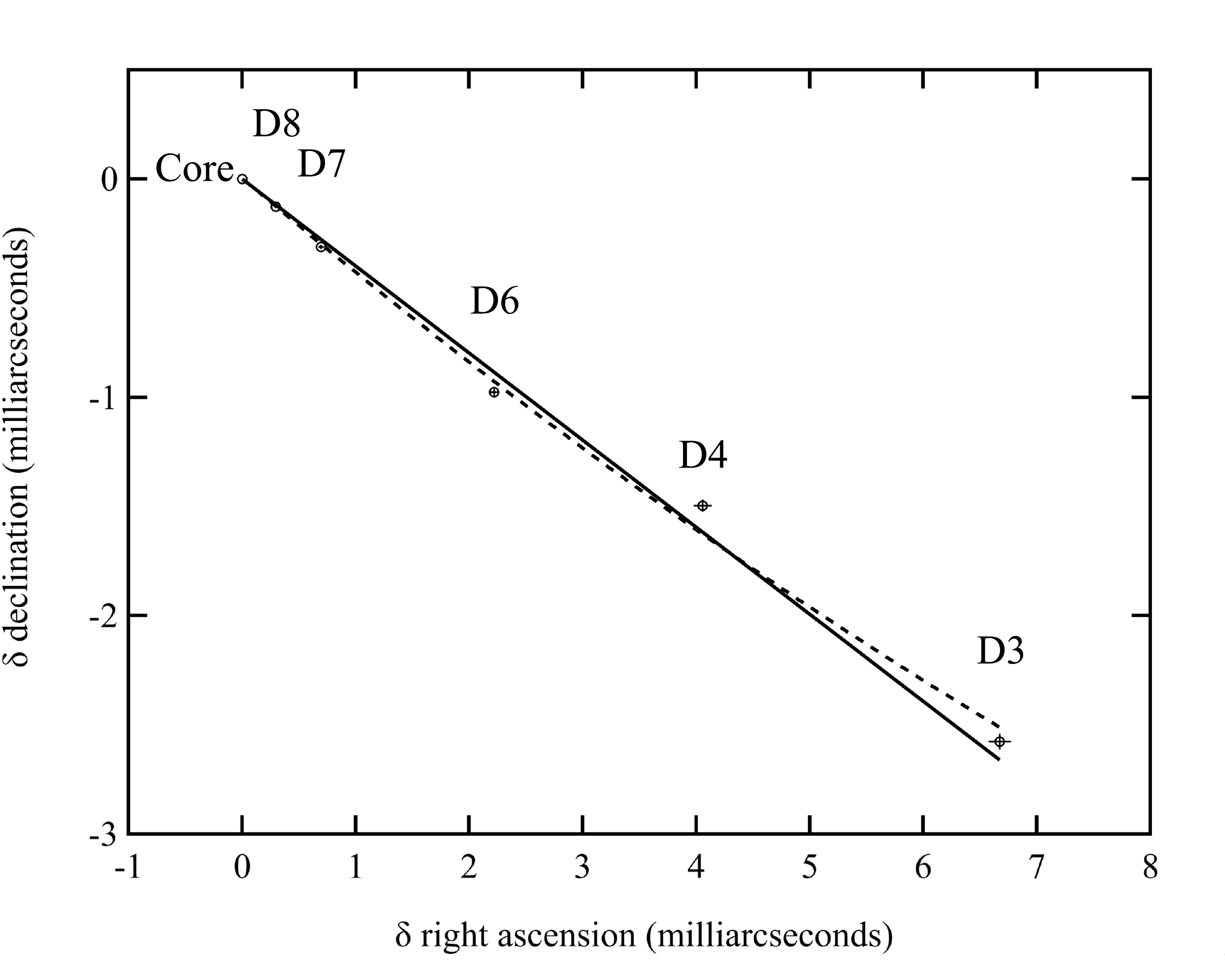}
\caption{The mean offsets from the radio core,  of components D8, D7, D6, D4 \& D3 in the jet of PKS 1413+135, averaged over the 23 epochs of the MOJAVE 15 GHz observations \citep{2019ApJ...874...43L}. Also shown are the error-weighted least-square linear (solid line)  and quadratic (dashed line) fits to the jet components. Note that the uncertainties in the mean offsets in both coordinates are much smaller than the size of the symbol for D8, comparable to the size of the symbol for D7 and D6, and slightly larger than the size of the symbol in D4 and D3. } 
\label{plt:collimation}
\end{figure}

We began this study by considering three SAV candidate events, in 1982, 1993, and 2000 in addition to the two, in 2009 and 2014, reported in Paper 1. But we found that the candidate SAV in 2000 (SAV3) is anomalous because it is not symmetric.  We nevertheless retained SAV3 in our analysis as an interesting example of intrinsic variability with achromatic similarities to the other four bona fide SAVs. Thus we attempt to model all 4 bona fide SAVs and in addition the anomalous SAV3 event as moving source components behind a single stationary, unvarying lens.

The purpose of this paper is twofold. First, to develop a robust nested sampling methodology that is able to simultaneously fit multiple lensing events in multiple bands, and second to test the gravitational millilensing hypothesis for all four of the bona fide SAVs that we have identified in  PKS 1413+135. 

In \textsection\ref{sec:observations} we discuss 40 years of multifrequency radio light curves from 1980 to 2020 and 24 epochs of VLBI observations from 1994 to 2011 of PKS 1413+135, which enable us to determine which features in the radio structure are definitely not lensed (the {\it unlensed} components) and which features might be lensed and therefore need to be {\it demagnified} in order to be able to determine what the PKS 1413+135 radio light curves would look like in the absence of the millilensing we are proposing; in \textsection\ref{sec:sample} we describe our gravitational lens fitting methodology;  in \textsection\ref{sec:res} we describe the results of the gravitational lens fitting modeling; in \S \ref{sec:jetmodels} we present three jet models and discuss their viability on the millilensing hypothesis; in \textsection\ref{sec:difficulties} we discuss two potential problems with the gravitational millilensing hypothesis;  in \textsection\ref{sec:lenses} we discuss possible millilenses and the recent potential discovery of a $\sim 10^4 M_\odot$ intergalactic gravitational lens through $\gamma-$ray burst (GRB) measurements by \citet{2021NatAs.tmp...55P}; in \S\ref{sec:periodicity} we explore possible SAV periodicity and in \textsection\ref{sec:disc} we discuss our findings.

For consistency with our previous papers, we assume the following cosmological parameters: $H_0 = 71$\,km\,s$^{-1}$\,Mpc$^{-1}$, $\Omega_{\rm m} = 0.27$, $\Omega_\Lambda = 0.73$ \citep{2009ApJS..180..330K}. None of the conclusion would be changed were we to adopt the model of \citet{2020A&A...641A...6P}.

\section{Observations}\label{sec:observations}

The radio observations that we used in our gravitational lensing analysis span  40 years and were made by the University of Michigan Radio Astronomy Observatory (UMRAO) at 4.8, 8, and 14.5~GHz, the Arc Minute Imager (AMI) of the Mullard Radio Astronomy Observatory (MRAO) at 15 GHz, the Owens Valley Radio Observatory (OVRO) at 15~GHz, the Mets\"ahovi Radio Observatory (MRO) at 22 and 37~GHz, the Institut de Radioastronomie Millimetrique (IRAM) at 90~GHz, and the Submillimeter Array (SMA) at 230~GHz and 340 GHz. Fig. \ref{plt:lightcurves} shows the flux density monitoring observations, with the shaded areas marking the bona fide SAVs and the cross-hatched shading showing the anomalous SAV3 event. In addition to these light curves, we used very long baseline array (VLBA) observations by \citet{1996AJ....111.1839P} and by the MOJAVE collaboration \citep{2019ApJ...874...43L}. By  good fortune the \citet{1996AJ....111.1839P} VLBI observations were made during SAV2, nine of the 23 epochs of the MOJAVE VLBI observations coincided with the anomalous SAV3 event, and another coincided with SAV4.

\begin{figure}[h!]
\includegraphics[width=0.4\textwidth]{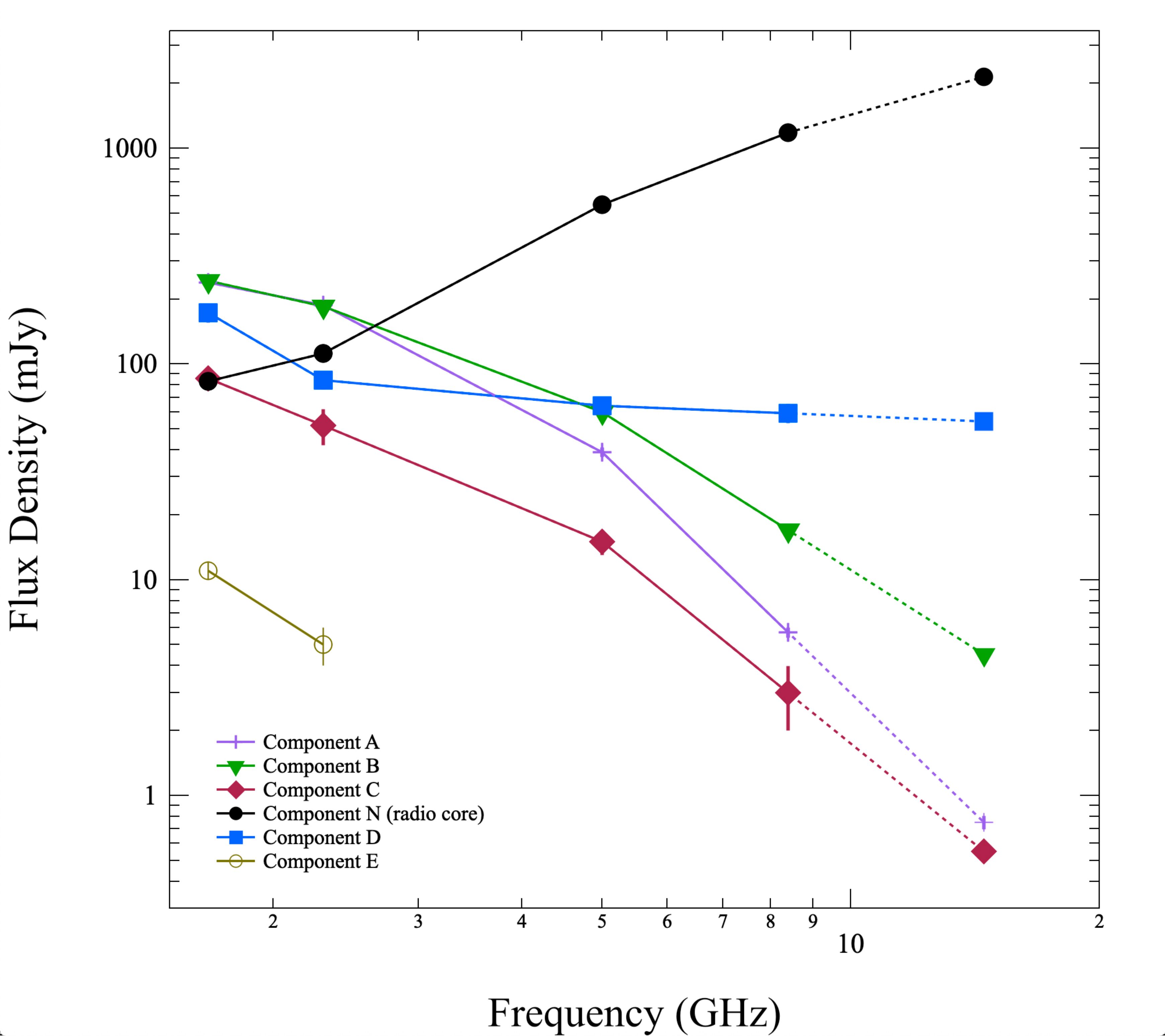}
\caption{Spectra taken from \citet{1996AJ....111.1839P}, based on  VLBA observations on 10-11 July 1994 (MJD 49543-4), and, for components A, B, C \& D, extrapolated to 14.5 GHz, as indicated by the dotted lines. The 14.5 GHz point for component N  is based on the total flux density measured by UMRAO minus the extrapolated values for the other components (see text).}
\label{plt:spectra}
\end{figure}

\subsection{Where Does SAV Originate in PKS 1413+135?}\label{sec:origin}
{We are interested in the most compact emission regions in the vicinity of the unresolved flat spectrum radio core.  The radio structure in this region at 15 GHz and 43 GHz is shown in Fig. \ref{plt:structure}.} The 15 GHz image is a stacked image made from MOJAVE maps at 23 epochs. {At this frequency, the core component ``N'' is by far the brightest and most compact feature, with an angular size (FWHM) of $0.16 \pm 0.02$ mas $\times \;\;0.08 \pm 0.01$ mas in PA $52^\circ \pm 14^\circ$.} The fraction of the total flux density in the core at 15 GHz determined from the 23 MOJAVE images is $0.73 \pm 0.14$, where the uncertainty is the standard deviation of the values. The brightest component outside the core is D8. The fraction of the flux density of D8 relative to the total flux density in the 23 MOJAVE images is $0.18 \pm 0.11$, where again the uncertainty is the standard deviation of the values. Given the fractional changes in total flux density during SAVs seen in Fig. \ref{plt:lightcurves}, it is clear that the components outside of the core are far too faint to be responsible for SAV, and that SAV features, whatever their origin, are due to changes in the core flux density.  For example, the MOJAVE image made from observations taken during SAV4 show that 81\% of the flux density is in the core, whereas only 9\% is in component D8. Similarly, these fractions in the images made from observations immediately preceding (following) SAV4 are 82\% vs. 12\% (71\% vs. 12\%). Since the morphology of PKS 1413+135  shows no evidence of the multiple images expected if there is gravitational lensing outside of the core  the components responsible for SAV must be located in the unresolved core.

\subsection{The opening angle, or cone angle, of the jet}\label{sec:opening}
Fig. \ref{plt:collimation} shows the measured offsets of the components of the PKS 1413+135 jet relative to the core. The error-weighted linear least-squares fit to the component positions is shown by the solid line, while that for a quadratic fit is shown by the dashed line. 
{The jet axis is defined as the 3D jet direction vector with respect to the core.
The position angles of the jet axis, measured North to East, are $248.3^\circ\pm0.2^\circ$ for the linear fit.} At the position of D3, the jet position angle, $\xi{\rm (D3)}$, in the quadratic fit is $\xi{\rm (D3)} =252.6^\circ\pm1.3^\circ$. The change in position for the curving jet between the core component and D3 is $\delta \xi{\rm (D3)} = 4.4^\circ\pm1.3^\circ$.

In PKS 1413+135 we clearly see a jet, which means that once outside the core, the line of sight lies outside the cone of the jet.  This means that, outside of the core, the angle between the jet axis and the line of sight, $\theta$, is greater than half the deprojected cone opening angle, $\zeta_{\rm dep}$. So we have $\theta>\zeta_{\rm dep}/2$.

For a straight jet, the observed cone angle, $\zeta_{\rm obs}$, does not vary with distance along the jet and can therefore  be measured at a single projected radial angular distance, $r$. 
The high-quality image resulting from the stacked MOJAVE measurements shown in Fig. \ref{plt:structure} (b) enables us to measure the opening angle of the jet in PKS 1413+135.
We measure a jet width of 2.75 mas at a distance of 7.5 mas from the core with a $1.84 \times 0.45 \; {\rm mas}^2$ beam having its major axis almost orthogonal to the jet axis. Deconvolving with the beam  gives a jet width of 2.04 mas and hence a cone angle of $15.6^\circ \pm 2^\circ$.  From the MOJAVE observations,  \citet{2017MNRAS.468.4992P} have estimated the mean width of the jet to be $11.1^\circ \pm 0.5^\circ$. This may be compared with the variation in position angles of the components, which is $12.4^\circ \pm 2^\circ$. Since these values are all in agreement within the errors, we will adopt the value of \citet{2017MNRAS.468.4992P} since this is the most accurate estimate.

Note that for $\theta \ll 1$ rad the deprojected cone angle of the jet is $\zeta_{\rm dep}=\zeta_{\rm obs} {\rm sin}(\theta)\sim \zeta_{\rm obs} \theta$, where the subscript ``$_{\rm dep}$''  denotes the deprojected jet when viewed at an angle of $\theta = 90\degr$ between the jet axis and the line of sight. An $11^\circ$  cone angle observed at angle $\theta$ to the jet axis has a deprojected cone angle of 0.19$\theta$, where here $\theta$ is in radians. Since, in the case of PKS 1413+135, we see a conical jet, it is clear that the lines of sight to components other than the core lie outside of the cone of the jet.

\begin{figure*}[ht!]
\centering
\includegraphics[width=1.0\textwidth]{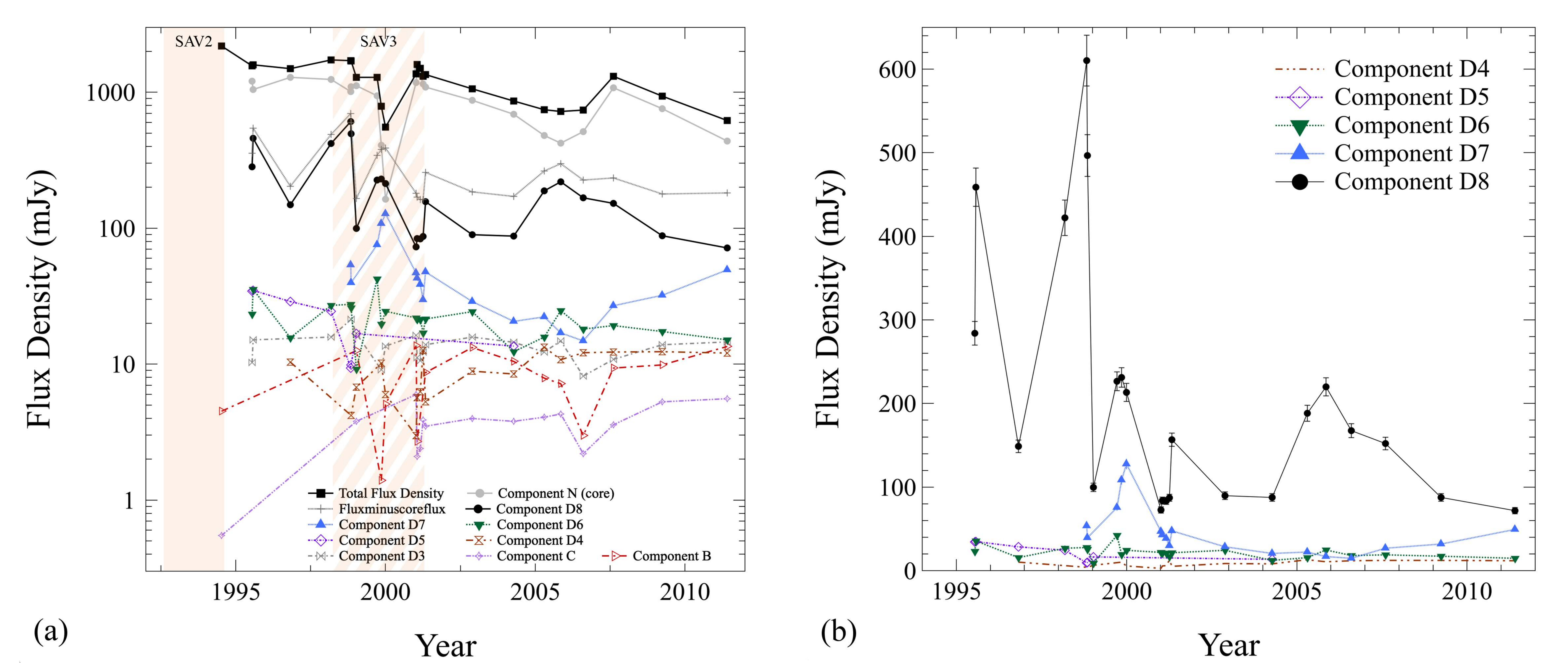}
\caption{PKS 1413+135 compact structure decomposition from VLBI observations at 24 epochs from 10 July 1994 to  26 May 2011.  The first VLBI observation shown is that of \citet{1996AJ....111.1839P}, which coincided with the peak in flux density towards the end of SAV2. The other observations are from MOJAVE \citep{2019ApJ...874...43L}. All errors in the MOJAVE  observations are $\pm 5 \%$.  Nine of the MOJAVE series of observations were made during SAV3, and these show clearly that SAV3 is not a symmetric variation after all. It has now been rejected as an SAV (see \S \ref{sec:sav3}). The two most rapid changes in the flux density of D8 shown in (b) provide estimates of the variability Doppler factor.} 
\label{plt:components}
\end{figure*}

\subsection{Multifrequency VLBI Observations}\label{sec:multifrequency}

The multifrequency VLBI observations of PKS 1413+135 by \citet{1996AJ....111.1839P} were made on 10-11 July 1994, i.e., on MJD 49543-4, at 1.67 GHz, 2.3 GHz, 5 GHz and 8.4 GHz. These observations coincided with the second peak of SAV2, at 14.5 GHz. In Fig. \ref{plt:spectra} we have plotted these data and extrapolated them to 14.5 GHz for Components A, B, C \& D.  UMRAO observations  made on MJD 49536 and 49551 gave 14.5 GHz flux densities of $2220 \pm 80$ mJy and $2180 \pm 110$ mJy, respectively. Thus the interpolated 14.5 GHz flux density on MJD 49543-4 is $2200 \pm 68$ mJy.  The sum of the flux densities of component A, B, C \& D is $60 \pm 7.5$ mJy.  Thus, the flux density of the radio core, component N, derived at 14.5 GHz, is $2140 \pm 68$ mJy. This is the value we have plotted in Fig. \ref{plt:spectra}. 

Note that the 14.5 GHz flux density of components outside the core, i.e. components A, B, C, D3, D4, and D6, amounted to  $60 \pm 7.5$ mJy, or $2.7 \%$ of the total flux density. Components D7 \& D8 were too close to the core to be resolved out in the \citet{1996AJ....111.1839P} observations. We return to this point  in \S \ref{sec:unlensed}.

\begin{deluxetable*}{l@{\hskip 8mm}cccccc}
\tablecaption{Variability Doppler Factor in PKS 1413+135\label{tab:xyz12}}
\tablehead{MJD Range&redshift&$r$&$\delta S_{\rm 15 GHz}$&$\delta t$&$T_{\rm var}$&$D_{\rm var}$\\
&&(Gpc) & (mJy)& (yr)& (K)&}
%\decimalcolnumbers
\startdata
49915-49928  &0.247&0.989 &$174.5 \pm 12.3$&0.036&$(6.7\pm 0.5)\times 10^{13}$&$10.8\pm0.8$\\ 
&0.5 &1.883 &&&$(2.4\pm 0.2)\times 10^{14} $&$20.0\pm1.4$\\ 
51116-51124 &0.247&0.989 &$113.4 \pm 8.0$&0.022&$(11.3\pm 0.8)\times 10^{13} $&$13.0\pm0.9$\\ 
&0.5&1.883 &&&($4.1\pm 0.3)\times 10^{14} $&$24.0\pm 1.7$\\ 
\enddata
\tablecomments{Variability Doppler factors derived from the variations in component 8 determined by \citet{2019ApJ...874...43L} for the limiting  redshifts of PKS 1413+135 determined in Paper 3.}
\label{tab:vardop}
\end{deluxetable*}

\begin{figure*}[ht!]
\centering
\includegraphics[width=1.0\textwidth]{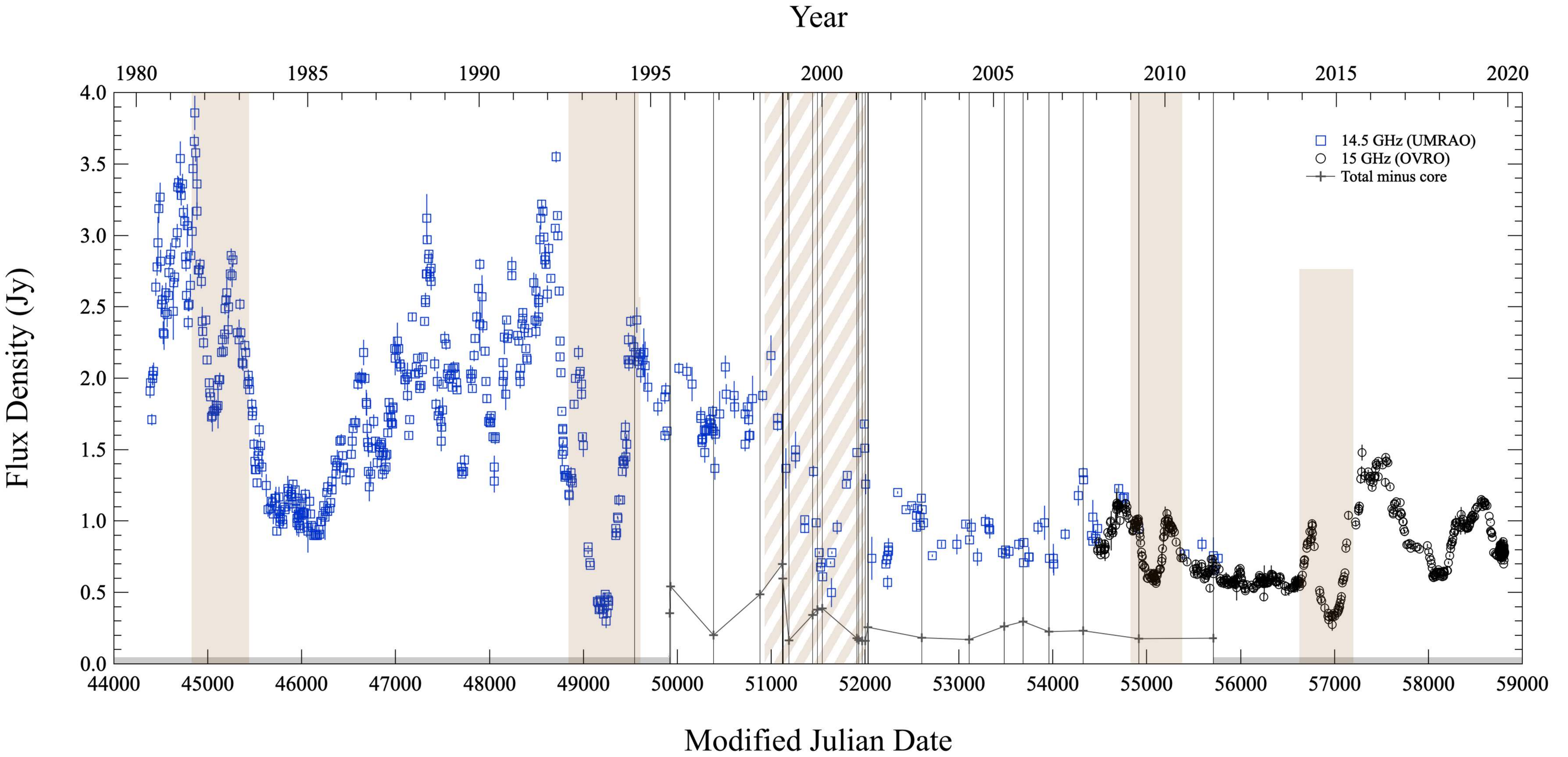}
\caption{The UMRAO 14.5 GHz (blue open squares) and OVRO 15 GHz (black circles) flux densities from 1980 to 2020.  The vertical filled shaded regions show the four bona fide  SAVs, and the cross-hatched region shows the anomalous SAV3 event, discussed in this paper. The vertical black lines indicate the epochs when VLBI observations discussed in the text were made. The gray crosses show the total flux density of the unlensed components from the MOJAVE observations \citep{2019ApJ...874...43L}. The gray bar shows the constraint ($44.4 \pm 1.9$ mJy) derived from the unlensed components in the MOJAVE observations at 15 GHz. A linear interpolation of the entire grey trace is subtracted from the total flux densities plotted here in the millilensing model fitting of SAVs.}
\label{plt:constraints}
\end{figure*}

\subsection{The Moving Components in PKS 1413+135}\label{sec:moving}

If SAV is caused by gravitational millilensing it can only be due to the core since this dominates the 15 GHz flux density.  Thus SAV should give rise to multiple stationary images of the radio core, but, as shown in the MOJAVE observations and the examples in Fig. \ref{plt:structure}, no stationary components, which could be multiple images of the radio core, are seen down to a separation of 250 $\mu {\rm as}$.  We therefore take 250 $\mu {\rm as}$ as an upper limit to the Einstein radius of the putative millilens.

The separations of the  components of PKS 1413+135 from the core are given by \citet{2019ApJ...874...43L}. The most recent component to emerge from the core as seen in the MOJAVE 15 GHz observations,  ``component 8'' (D8), is more than 250 microarcseconds from the core and in view of both its motion and its distance from the core it is clearly not lensed. This component is not distinguishable from the core in VLBI observations below 15 GHz, and hence was not resolved in the observations of  \citet{1996AJ....111.1839P}. {This might appear to be a rather conservative estimate of the Einstein radius, and one might wonder whether the core shift might be measured between epochs. We have looked at the possibility of measuring core shifts in the 15 GHz MOJAVE data sets,
and we find that with the VLBA resolution at this frequency and the data quality, it would require a core
shift of 0.2 mas to be able to measure it with any confidence. One might also wonder why we do not take half of this value to be the Einstein radius. The answer can be seen in Appendix A of Paper 3.  Although for a lens perfectly aligned with  the background source the separation of the two images is indeed twice the Einstein radius, the separation rapidly approaches the Einstein radius for impact parameters larger than the Einstein radius.}

  \subsubsection{The Variability Doppler Factor in PKS 1413+135}\label{sec:Doppler}

The flux density variations of the pc-scale components detected in the MOJAVE observations \citet{2019ApJ...874...43L} are shown in Fig. \ref{plt:components}. It can be seen here that the 15 GHz flux density of the component closest to the core (D8) varied rapidly between 1995 and 1998. The most extreme variations occurred in the two periods MJD 49915-49928, and MJD 51116-51124.  The variations and timescales are given in Table \ref{tab:vardop}.

We can calculate the variability brightness temperature from equation B14 of Paper 3. To do this we need the comoving coordinate distance, $r$, of PKS 1413+135. We showed in Paper 3 that PKS 1413+135 has redshift $0.247< z < 0.5$, which yields  $988.8< r < 1883.1$ Mpc. The corresponding variability brightness temperatures, $T_{\rm var}$, and variability Doppler factors, $D_{\rm var}$, are given in Table \ref{tab:vardop}, where we have assumed that the emission frame brightness temperature is the equipartition brightness temperature, which  $\sim 10^{11}$ K  \citep{1994ApJ...426...51R,2018ApJ...866..137L}.  We see from Table \ref{tab:vardop} that $D_{\rm var} = 10.8 \pm 0.8 \rightarrow 13.0 \pm 0.9$ for z=0.247, and $D_{\rm var} = 20.0 \pm 1.4 \rightarrow 24.0 \pm 1.7$ for z=0.5.  These values agree with the values presented in Table 4 of Paper 3, which were determined from the UMRAO total flux density light curves.

\subsection{The Unlensed Components of PKS 1413+135}\label{sec:unlensed}

The jet components that have been detected at separations from the radio core greater than 250 $\mu {\rm as}$ are all moving. None of them is stationary relative to the core, and it is therefore clear that these components are unlensed.
We distinguish between these unlensed components and components buried in the core that may  be lensed and for which we estimate their demagnified flux densities.  

The components detected in the jet and counterjet of PKS 1413+135 are all moving relatively slowly for a blazar, and therefore no dramatic structural changes have occurred over the period of observations we are considering. While it is possible that one of the counterjet components ``C'' seen in Fig. \ref{plt:structure} (b) is a multiple image of the core, we reject this possibility because of the extremely good alignment of these components with the core ``N'' and component ``B''.

In view of the rapid increase in the flux density of component D8 between July 17 and July 30 1995, it is by no means clear what the flux density of D8 was before July 1995 and hence during SAV2. We also note that component D7 was first  detected in 1998. For these reasons we do not include components D7 and D8 as constraints in our  gravitational lens fitting of SAV2 and SAV5, but we do use the flux density of the other unlensed components discussed in \S \ref{sec:multifrequency} as a constraint.   For SAV2 and SAV5 the estimate of the flux densities of the components, apart from the core, D8 and D7,  from the  VLBI observations of \citet{2019ApJ...874...43L}  is $S_{\rm 15 GHz} = 44.4 \pm 1.9$ mJy.

In fig. \ref{plt:constraints} we show the UMRAO 14.5 GHz and OVRO 15 GHz light curves.
{Note that the source is much more variable prior to 2008.
Variability in the form of flaring is common in blazars and generally attributed to the heating and subsequent cooling of localized regions within the jet where particles are accelerated. The timescales of individual flares can range from days to months or longer, and the behavior is generally stochastic with no pattern in temporal spacing or maximum flux amplitude. It was fortunate that the stochastic behavior in PKS 1413+135 decreased around 2008, in time for the SAV events of 2009 and 2014 to be clearly visible.}
In fig. \ref{plt:constraints}, 
gray crosses show the flux density level due to components outside the core as measured by MOJAVE \citep{2019ApJ...874...43L}. The unlensed flux density level used as a constraint for SAV2 and SAV5 is shown by the gray bar in Fig. \ref{plt:constraints}. For the remaining observed frequencies, we conservatively estimate the spectral indices for each unlensed component from fig.~\ref{plt:spectra} and calculate the total unlensed flux dneisty for each SAV using the known 14.5 -- 15GHz component flux densities. These unlensed components are subtracted before any millilens fitting.

\begin{figure}[h!]
\includegraphics[width=0.5\textwidth]{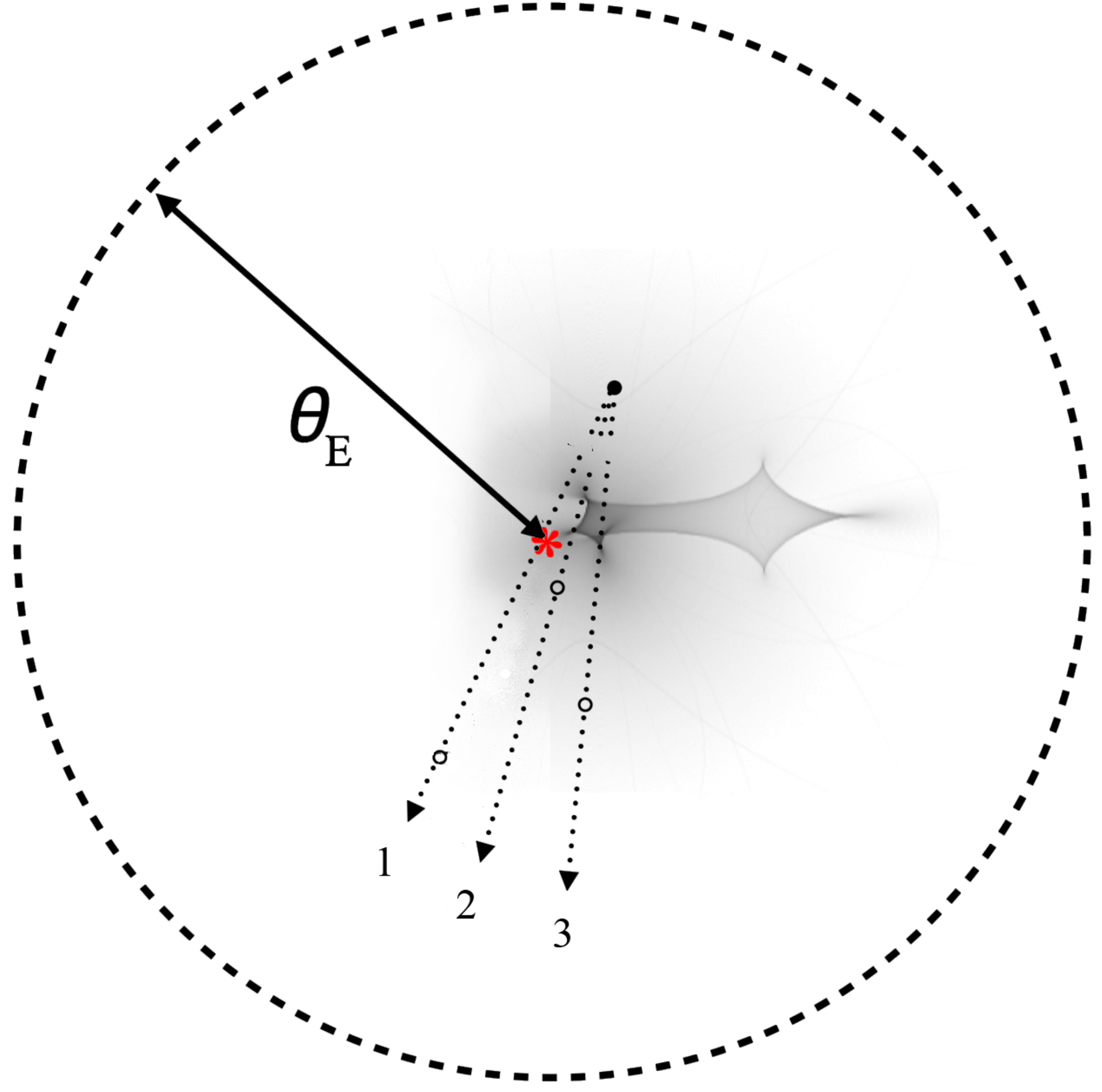}
\caption{Schematic of the jet and millilens. Component trajectories from the central engine (black dot) pass behind the caustic pattern (grayscale) produced by a binary lens. The red asterisk marks the position of the larger mass. The Einstein radius is  indicated by the dashed circle centered on the larger mass. The smaller mass is beyond  the Einstein radius in PA $-90^\circ$.  The open circles represent components that are $\sim 1\%$ of the size of the Einstein radius.  We show  in  \S \ref{sec:res} that trajectories close to 2 produce SAVs. For an apparent jet cone angle $\zeta_{\rm app}\sim 10^\circ$ we would see SAV from components moving  between trajectories 1 and 2.}
\label{plt:cartoon}
\end{figure}

 \subsection{The Anomalous SAV3}\label{sec:sav3}

In Paper 1 the authors suggested  SAV3 as a candidate lensing event, and as can be seen in Fig. 2 of Paper 1, it does look like a good SAV candidate.  However they did not at the time have the UMRAO 8 GHz and 14.5 GHz lightcurves, nor did they have the MOJAVE VLBI results. As can be seen in Fig. \ref{plt:lightcurves}, the UMRAO 8 GHz and 14.5 GHZ lightcurves show that the symmetry is not good.  The MOJAVE 15 GHz results also show this very clearly, as can be seen in Fig. \ref{plt:components} (a). ``SAV3'' is not a symmetric variation after all. It is therefore of interest that, as we will see in \S \ref{sec:res}, we have been unable to fit SAV3 well with the same gravitational lensing model that fits all four of the other SAVs. It is likewise of interest that SAV3 does not fit the period of 989 days that fits the other four SAVs, as discussed in \S \ref{sec:periodicity}.

\section{Gravitational lens Fits}\label{sec:sample}

Using SAVs 4 and 5, the authors showed in Paper 1 that if SAVs are caused by gravitational lensing, the Einstein radius $\theta_E \lesssim 250\mu {\rm as}$ and the angular source size $\sigma_s$ must be small compared to $\theta_E$: $\sigma_s / \theta_E \lesssim 0.03$.

Since multiple imaging is not observed in this source, neither time delay nor relative image brightness can be used to analyse the possible lensing system. We must then use a millilensing forward model of the observed magnifications. We develop a light curve fitting procedure that can simultaneously fit all observed frequencies of an SAV (or multiple SAVs) to find the optimal lens parameters. We will use this procedure to show that the four bona fide SAVs are consistent with a single lensing model.

\begin{figure*}[ht!]
\centering
\includegraphics[width=0.7\textwidth]{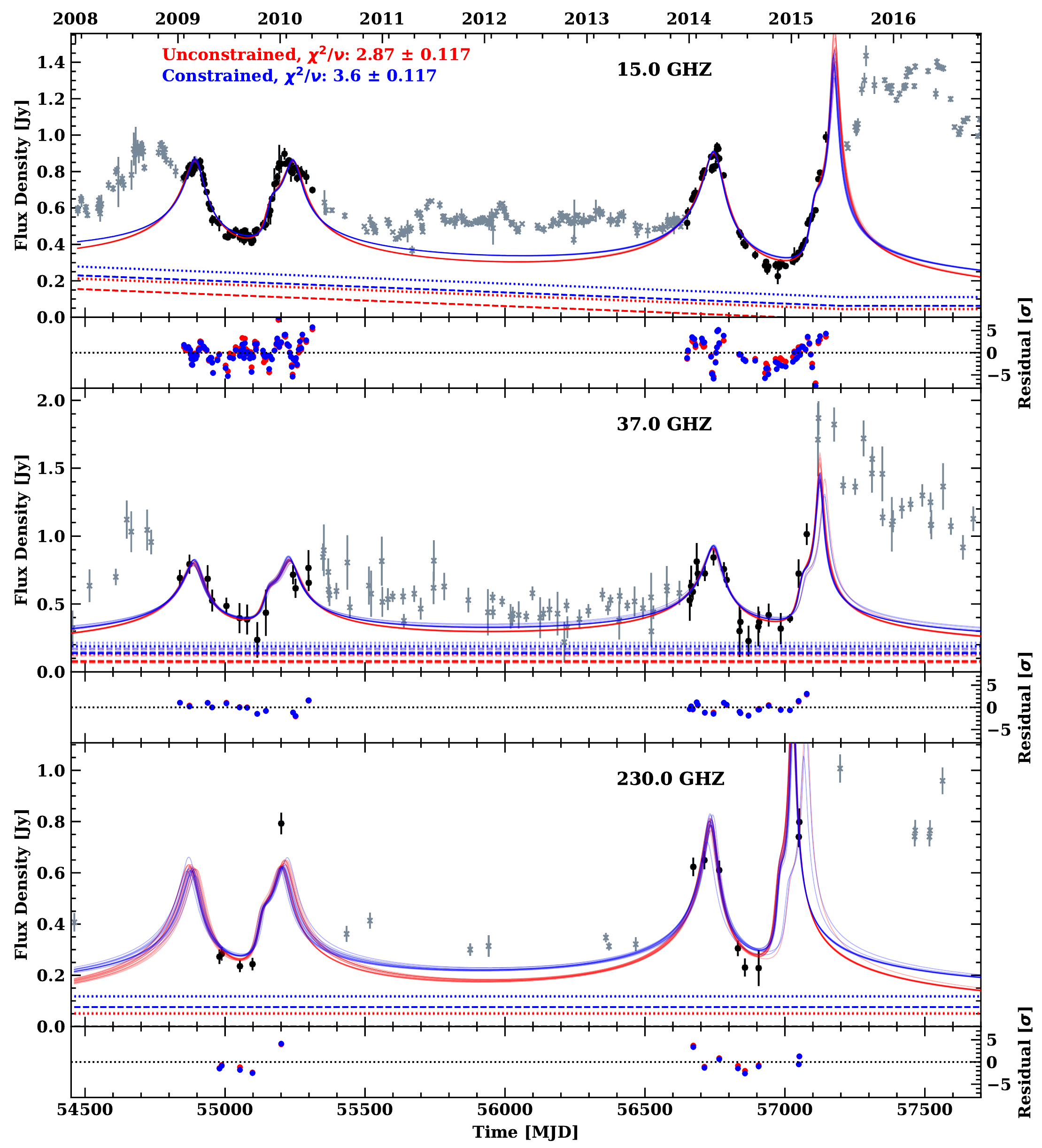}
\caption{SAV4 and SAV5 fit results using the detrended 15~GHz, 37~GHz and 230~GHz core light curves for the simultaneous fit of SAV1,2,4,5. The sum of both SAV magnifications is fitted to the data. The SAVs are defined by the black points. Model residuals in units of standard deviation from the mean are shown below each panel.
The blue fit assumes the core unlensed flux density must be $ \geq 1/3$ of the minimum SAV flux density.
Dashed traces represent a posterior samples of the total core unlensed flux density level.
Dotted traces show the total core demagnified flux density, i.e. the core flux density that would be observed if there were no magnification. 
Each fit shows multiple posterior samples. Chi-squared per degree of freedom values include all frequencies. Best fit parameters for the blue (constrained) fit are shown in Table~\ref{tab:val} and posteriors in fig.~\ref{plt:corn}.}
% Blue shows multiple posterior samples for the individual lensing events; their (fitted) sum is shown in red. Best fit parameters are shown in table~\ref{tab:val} and posterior distributions in the appendix.}
\label{plt:sav45}
\end{figure*}

\begin{figure}[ht!]
\includegraphics[width=0.5\textwidth]{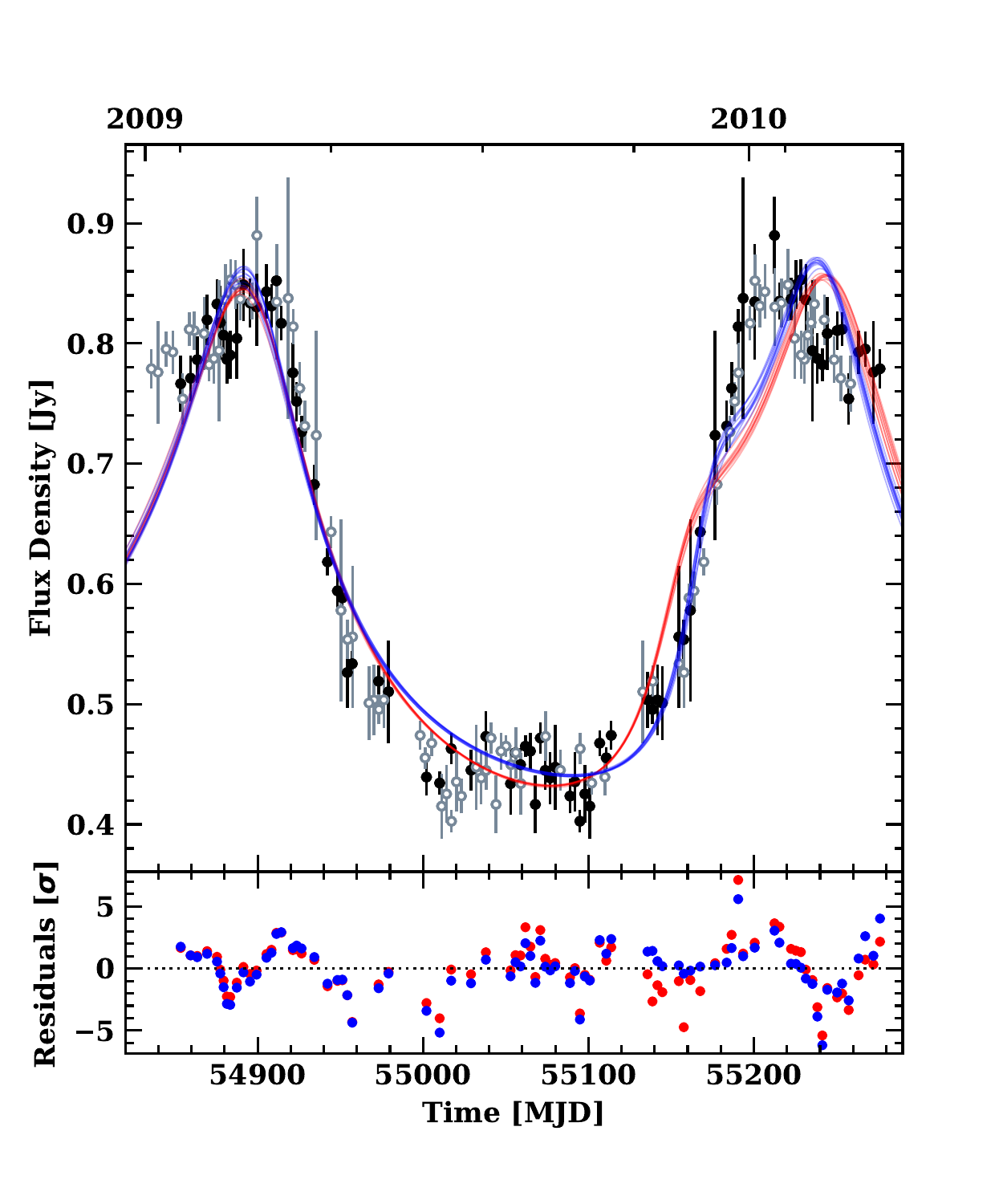}
\caption{The symmetry in SAV4, which is the best example we have of SAV. The filled circles are the original data and the open circles show this data reflected about the symmetry axis. Also shown is the quality of the fits of the models to SAV4. The curves are as described in Fig. \ref{plt:sav45}. Model residuals in units of standard deviation from the mean are given in the pull plot below.}
% Blue shows multiple posterior samples for the individual lensing events; their (fitted) sum is shown in red. Best fit parameters are shown in table~\ref{tab:val} and posterior distributions in the appendix.}
\label{plt:symmetry}
\end{figure}

\begin{figure*}[]
\centering
\includegraphics[width=0.8\textwidth]{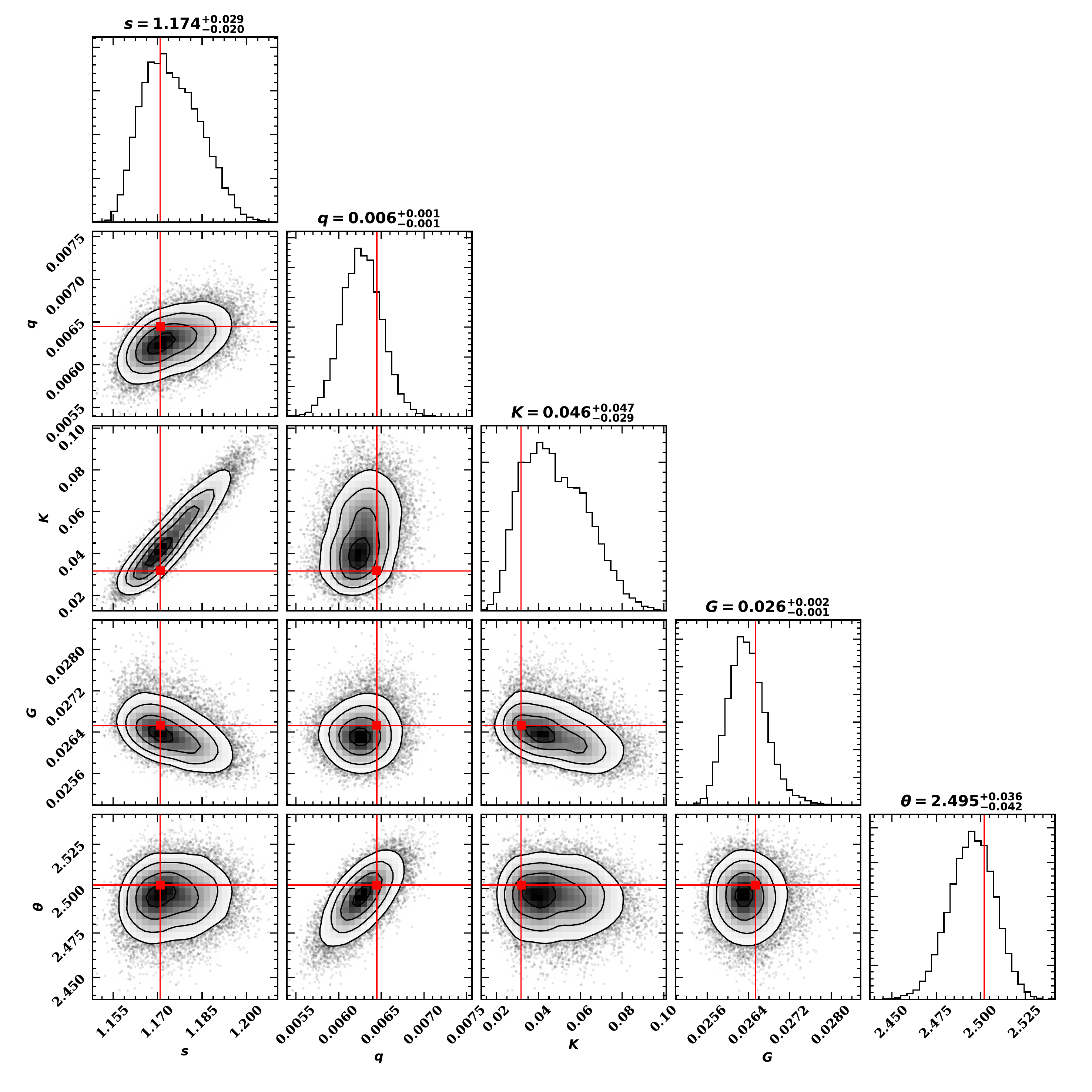}
\caption{SAV1,2,4,5 joint fit posterior distribution for intrinsic lens parameters. Red lines show maximum a posteriori values. $3\sigma$ confidence intervals are shown above the panels.}
\label{plt:corn}
\end{figure*}

\begin{figure*}
\resizebox{\hsize}{!}{\includegraphics[scale=1]{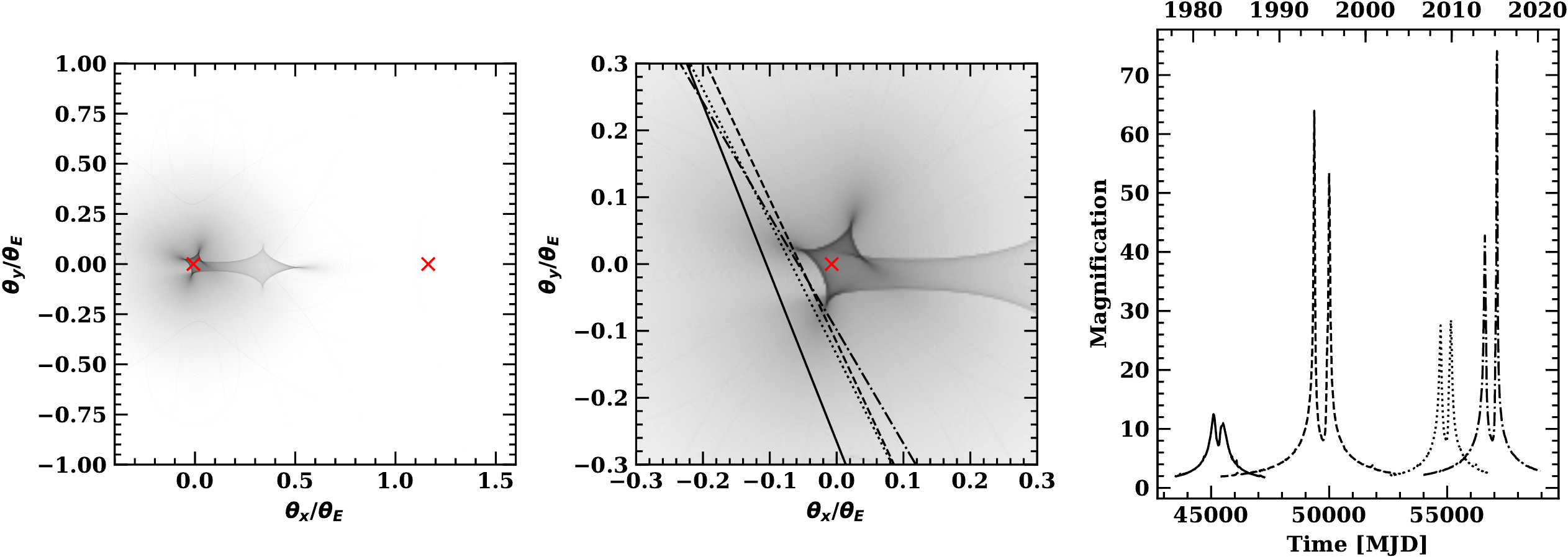} }
\caption{\textit{Left:} Caustics for the best fit binary lens model with constant external convergence and shear for SAV1,2,4,5 joint fit. \textit{Center:} Zoom of the caustics with the black lines showing the path of the source behind the lens plane (moving downwards) for the 4 individual SAVs, in units of the $\theta_E$. Red crosses show the projected locations of the binary masses, with the larger mass on the left in the left-hand figure. {The source tracks relative to the caustics have a low probability in general, and we use this to rule out  models where the source tracks are unconstrained -- for example models where the  line of sight lies within the jet cone -- see Fig. \ref{plt:Cartoonsav}(c).}  \textit{Right:} Magnification patterns caused by the source paths through the lens caustics for each individual SAV.}
\label{plt:caustic} 
\end{figure*}

As discussed in detail in Paper 1, before fitting SAVs, a specific lens model must be chosen. A simple elliptical mass distribution will produce a fast rise slow decline followed by slow rise fast decline (FRSD-SRFD), or ``crater'', profile in a light curve. However both SAV4 and SAV5 have slow rise fast decline followed by fast rise slow decline (SRFD-FRSD), or ``volcano'', profiles and thus require a more complex mass distribution than a simple elliptical.  
{The simplest model with sufficient caustic complexity to reproduce a ``volcano'' profile is a binary lens: a lensing system consisting of two point masses. For this reason we use a binary lens as our lensing model. }

As in Paper~1, we also include constant convergence and shear terms in our lens model, to account for the extended environment around the binary lens (for example the possible spiral host galaxy). Unlike Paper~1, we allow the constant shear to lie along any direction - this introduces an additional parameter $\phi$ ($0 \leq \phi < 2\pi$): the angle between the binary axis and the external shear ($\phi/2$). A schematic of the source-lens configuration is shown in Fig. \ref{plt:cartoon}. %These are constrained to be small enough that they are subdominant. 
Since the source is constrained to be small compared to $\theta_E$, we assume a point source. For a single lensing event, this model has 9 free parameters that must be constrained through fitting: 
\begin{itemize}
\item $t_0$ (time of closest source approach to the centre of mass of the lens system, in days)
\item $u_0$ (source impact parameter, in units of $\theta_E$)
\item $t_E$ (time taken to cross 2$\theta_E$, in days)
\item  $s$ (distance between the two point masses in the binary, in units of $\theta_E$)
\item $q$ (binary mass ratio)
\item $\alpha$ (angle relative to axis of binary lens of source path on the plane of the sky)
\item $k$ (external convergence, $k \geq 0$)
\item $\gamma$ (external shear, $\gamma \geq 0$)
\item $\phi/2$ (angle between binary axis and external shear)
\end{itemize}

Only $s, q, k, \gamma, \phi$ are intrinsic to the lens and should not change significantly between SAVs. 

\begin{table*}
\centering
    \footnotesize
    \begin{ttabular}
        \bfseries & $\mathbf{s}$ [$\theta_E$]& $\mathbf{q}$ & $\mathbf{k}$ & $\boldsymbol{\gamma}$ & $\boldsymbol{\phi}$ [rad] & $\alpha$ [rad] & $t_0$ [MJD] & $u_0$ [$\theta_E$] & $t_E$ [yrs] & $\eta_{0}$ & $\eta_{E}$ \\
        \bfseries SAV1&[$0.2,1.8$]&[$0,0.5$]&[$0,0.2$]&[$0,0.1$]&[$0,2\pi$]&[$0,\pi$]&[$44500,45650$]&[$-0.5,0.5$]&[$4,16$]&[$-0.05,0.05$]&[$-0.5,0.5$]\\
        \bfseries SAV2& -- & -- & -- & -- & -- & [$0,\pi$] &[$48500,49650$]&[$-0.5,0.5$]&[$4,16$]&[$-0.05,0.05$]&[$-0.5,0.5$]\\
        \bfseries SAV3& -- & -- & -- & -- & -- & [$0,\pi$] &[$52500,53650$]&[$-0.5,0.5$]&[$4,16$]&[$-0.05,0.05$]&[$-0.5,0.5$]\\
        \bfseries SAV4& -- & -- & -- & -- & -- & [$0,\pi$] &[$54500,55650$]&[$-0.5,0.5$]&[$4,16$]&[$-0.05,0.05$]&[$-0.5,0.5$]\\
        \bfseries SAV5& -- & -- & -- & -- & -- & [$0,\pi$] &[$56500,57650$]&[$-0.5,0.5$]&[$4,16$]&[$-0.05,0.05$]&[$-0.5,0.5$]\\
    \end{ttabular}
    \caption{Lens model uniform prior ranges for each SAV. Intrinsic lens parameters are in bold, these are shared by all SAVs during fitting.}
    \label{tab:prior}
\end{table*}

A millilensed light curve contains  no information about the orientation of the source on the sky and thus places no  constraints on the lens binary axis orientation, and hence on position angles on the sky of the trajectories of lensed components. But once fixed with respect to one SAV it fixes the orientation of the lens axis  relative to the trajectory of that particular component. Therefore, since the axis of the lens is fixed from one SAV to another, any changes in $\alpha$ reflect changes in the orientation of the trajectories of the different lensed components.  Since SAV4 is by far the best example we have of  an SAV we choose it as the fiducial SAV, and measure changes in $\alpha$ relative to its value for SAV4.

As shown in Paper 3, the jet axis in PKS 1413+135 is aligned within a few degrees of the line of sight. Thus, small fluctuations in the jet direction can cause large changes in the path of jet features across the sky. For this reason we allow $\delta\alpha$ to vary freely between SAVs. We find that the posterior $\delta\alpha$s are very small in all the observed SAV1, SAV2, SAV3 and SAV5 events.  This has important implications for the jet collimation in PKS 1413+135. 
%The value for SAV3 is significantly larger than for the other SAVs.

We expect some chromaticity in SAVs due to variation of the source size and centroid position with frequency, common in blazars \citep{1979ApJ...232...34B}. 
In some SAVs, for example SAV2, there is a visible trend of lower frequency lensing patterns occuring later in time, and persisting for a longer period of time. We allow both of these degrees of freedom for all SAVs in a simple way by parameterizing $t_0$ and $t_E$ as frequency $\nu$ dependent power laws. This includes two additional fitting parameters, power law spectral indices $\eta_0$ and $\eta_E$:
\begin{itemize}
\item $\eta_0$ (where $t_0(\nu) \propto t_0\nu^{\eta_0}$)
\item $\eta_E$ (where $t_E(\nu) \propto t_E\nu^{\eta_E}$)
\end{itemize}
where $\nu$ is measured in GHz. We find that the fits are vastly improved with these parameters included, \textsection\ref{sec:res}.

For a binary lens, the magnification and image positions can be calculated analytically for each point in the source's path following \citet{1995ApJ...447L.105W}. This involves solving a 5th order complex polynomial numerically. We include the constant convergence and shear, raising the polynomial to 9th order. To calculate magnification curves from our lens model, we augment the existing open source microlensing package \textit{MuLensModel} \citep{poleski_modeling_2019} (this makes use of \textit{VBBL} \citet{bozza_microlensing_2010} for fast polynomial solving and magnification calculations), to include a constant external convergence and shear. This augmentation is publicly available.

Since the binary lens model is highly non-linear and has a significant number of parameters, in order to fit the generated magnification curves to SAVs we turn to Bayesian nested sampling \citep{skilling_nested_2006} (specifically the MultiNest algorithm developed by \citet{feroz_importance_2019}). This produces joint posterior distributions over each of the final parameters and additionally calculates the model evidence (marginal likelihood), allowing direct comparison between different models.
We augment the standard importance nested sampling approach by solving a quadratic program (QP) to find the linear parameters of the lensing model. These are the linear flux density scaling and offset of the magnification curve to fit the data. Since these parameters are linear, they have  well defined optimal solutions using least-squares, and so it is much more efficient to solve for them directly rather than to include them as additional parameters in the nested sampling. However, one cannot simply use a least-squares approach because the parameters are constrained (the flux density scaling and offset must be positive). A constrained least-squares can be formulated as a QP \citep{boyd_van}.
QPs can be solved very quickly using the convex optimization package OSQP \citep{osqp}.

Our fitting procedure consists of running MultiNest using an augmented MuLensModel to produce the magnification curves for each parameter sample, calculating the likelihood for each sample (and the linear parameters) by solving a constrained least-squares problem.
The novelty of the method is in calculating the magnification curves quickly enough for nested sampling to work, this would not be possible using standard ray tracing techniques as we did in Paper 1, and reducing the parameter space by solving for the constrained linear parameters directly. Our method can be applied to multiple (different frequency) light curves simultaneously, each contributing to the likelihood for a specific parameter sample and each with their own linear flux density scaling and offset. The importance of each light curve can be weighted. In this work we treat each observation equally; frequencies with fewer observations and higher uncertainties contribute less to the fitting. %we don't use frequency importance weighting in this work. %light curve importance automatically "weighted" by the number of data points

Multiple lensing SAVs can be fitted simultaneously, with their summed flux density contributions fitted to the light curve. In the constrained least-squares framework, additional constraints can also be placed on the maximum or minimum flux density allowed; for our fits, we make the constraint that the total flux density should not go below zero anywhere. We also include a small regularization term penalizing non-smooth magnification curves in our fits. This avoids unlikely jagged and sharply peaked magnifications in the unsampled areas of the light curves and helps the fitting procedure converge more quickly.

Using our method, we are now able directly to fit SAVs. We have validated our lens model by reproducing the results from ray-tracing simulations used in Paper~1 and a number of other single and binary lens test cases.

\begin{table*}
\centering
\footnotesize
    \begin{ttabular}
        \bfseries & $t_0$ [MJD] & $\delta u_0$ [$\theta_E$] & $t_E$ [days] & $\mathbf{s}$ & $\mathbf{q}$ & $\delta\alpha[^{\circ}]$  & $\mathbf{k}$ & $\boldsymbol{\gamma}$ & $\boldsymbol{\phi}$ [rad] & $\eta_{0}$ & $\eta_{E}$ \\
        \bfseries SAV1&$45112.494^{+43.143}_{-70.513}$&$-0.026^{+0.004}_{-0.004}$&$3377.133^{+166.034}_{-230.442}$& -- & -- &$-5.006^{+1.110}_{-1.120}$& -- & -- & -- &$-0.0006^{+0.0003}_{-0.0003}$&$-0.124^{+0.053}_{-0.026}$  \\
        \bfseries SAV2&$49298.693^{+87.800}_{-153.205}$&$0.008^{+0.001}_{-0.001}$&$6535.783^{+64.019}_{-334.842}$&--&--&$-1.519^{+0.754}_{-0.851}$&--&--&--&$-0.0012^{+0.0005}_{-0.0003}$&$-0.193^{+0.031}_{-0.033}$\\
        \bfseries SAV3 & $51785.8411^{+77.9544}_{-202.4781}$&$-0.047^{+0.0585}_{-0.0184}$&$6551.6575^{+48.2343}_{-305.3139}$&$1.1641^{+0.0324}_{-0.0186}$&$0.0087^{+0.0013}_{-0.0013}$&$1.471^{+1.2125}_{-1.2833}$&$0.0188^{+0.0543}_{-0.0162}$&$0.0327^{+0.0026}_{-0.0019}$&$2.4303^{+0.0489}_{-0.0577}$&$-0.0009^{+0.0008}_{-0.0006}$&$0.0163^{+0.0037}_{-0.0232}$\\
        \bfseries SAV4 & $55096.679^{+49.589}_{-79.878}$&$-0.028^{+0.051}_{-0.030}$&$3267.599^{+208.333}_{-264.855}$&$1.174^{+0.029}_{-0.020}$&$0.006^{+0.001}_{-0.001}$&$296.644^{+0.889}_{-0.945}$&$0.046^{+0.047}_{-0.029}$&$0.026^{+0.002}_{-0.001}$&$2.495^{+0.036}_{-0.042}$&$-0.0002^{+0.0001}_{-0.0001}$&$-0.023^{+0.056}_{-0.047}$\\
        \bfseries SAV5&$57029.745^{+84.722}_{-137.009}$&$0.007^{+0.002}_{-0.002}$&$4936.219^{+207.969}_{-492.406}$&--&--&$3.709^{+0.850}_{-0.931}$&--&--&--&$-0.0005^{+0.0003}_{-0.0002}$&$-0.134^{+0.063}_{-0.015}$\\
    \end{ttabular}
    \caption{Lens model maximum a posteriori values for each SAV. $99.7\%$ ($3\sigma$) confidence intervals are shown alongside. Intrinsic lens parameters are in bold; these are shared by all SAVs except SAV3 that was fitted separately. $t_0$ and $t_E$ are quoted for 4.8GHz, the lowest available frequency; for other frequencies $t_E(\nu) = t_E(\nu / 4.8)^{\eta_E}$ (and similarly for $t_0$) where $\nu$ has units of GHz. $\delta u_0$ and $\delta \alpha$ are the differences in $u_0$ and $\alpha$ from the SAV4 value.}
    \label{tab:val}
\end{table*}

\begin{figure}
\includegraphics[width=0.5\textwidth]{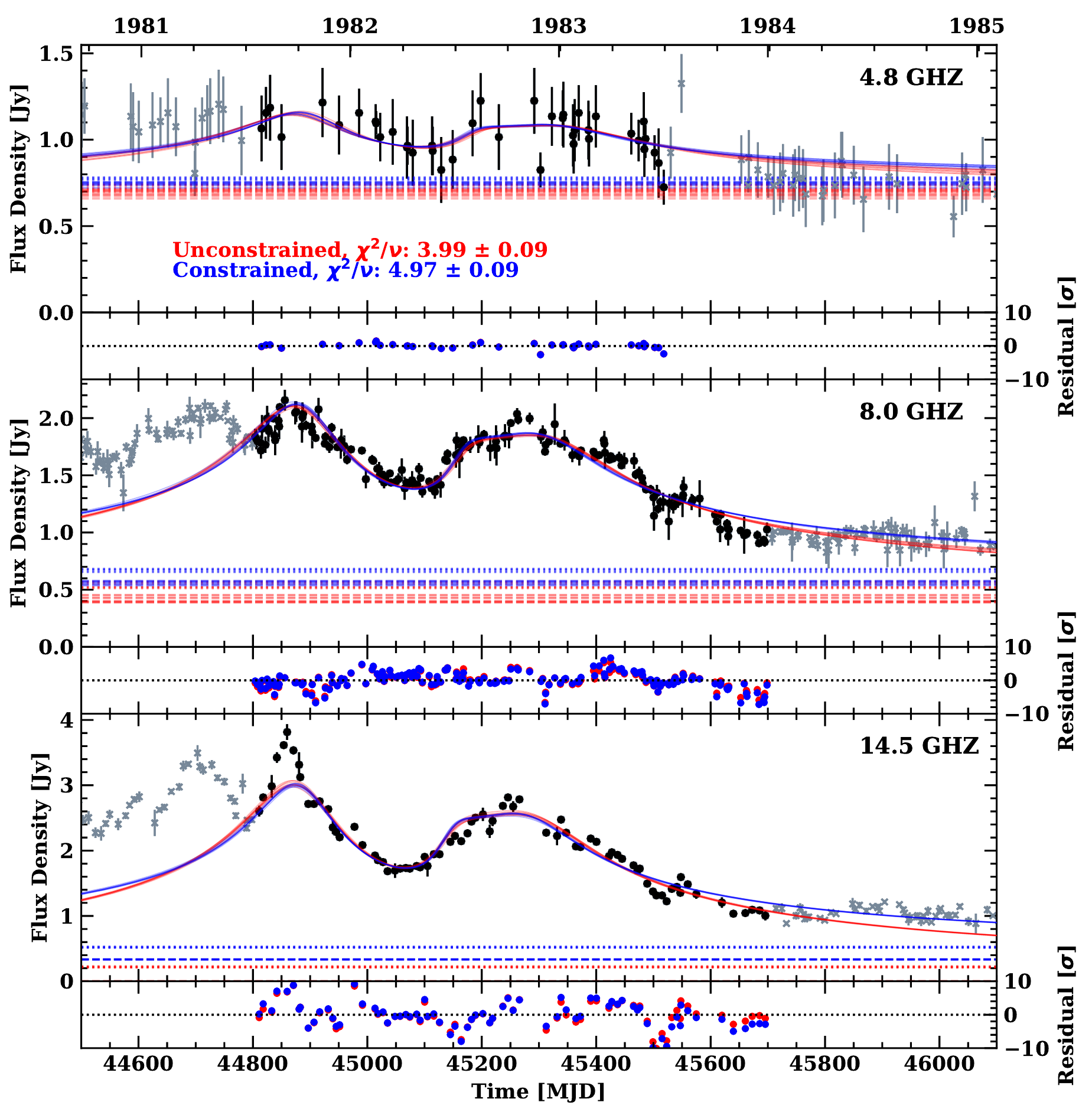}
\caption{Lensing model fits to SAV1, with colored traces as in fig.~\ref{plt:sav45}. All frequencies are fitted simultaneously, only the black points are used in the fitting. Red and blue lines are samples from their respective model posterior distributions. Best fit parameters are shown in table~\ref{tab:val}. No detrending is applied. Model residuals in units of standard deviation from the mean are shown below each panel.}
\label{plt:sav1}
\end{figure}

\begin{figure}
\includegraphics[width=0.5\textwidth]{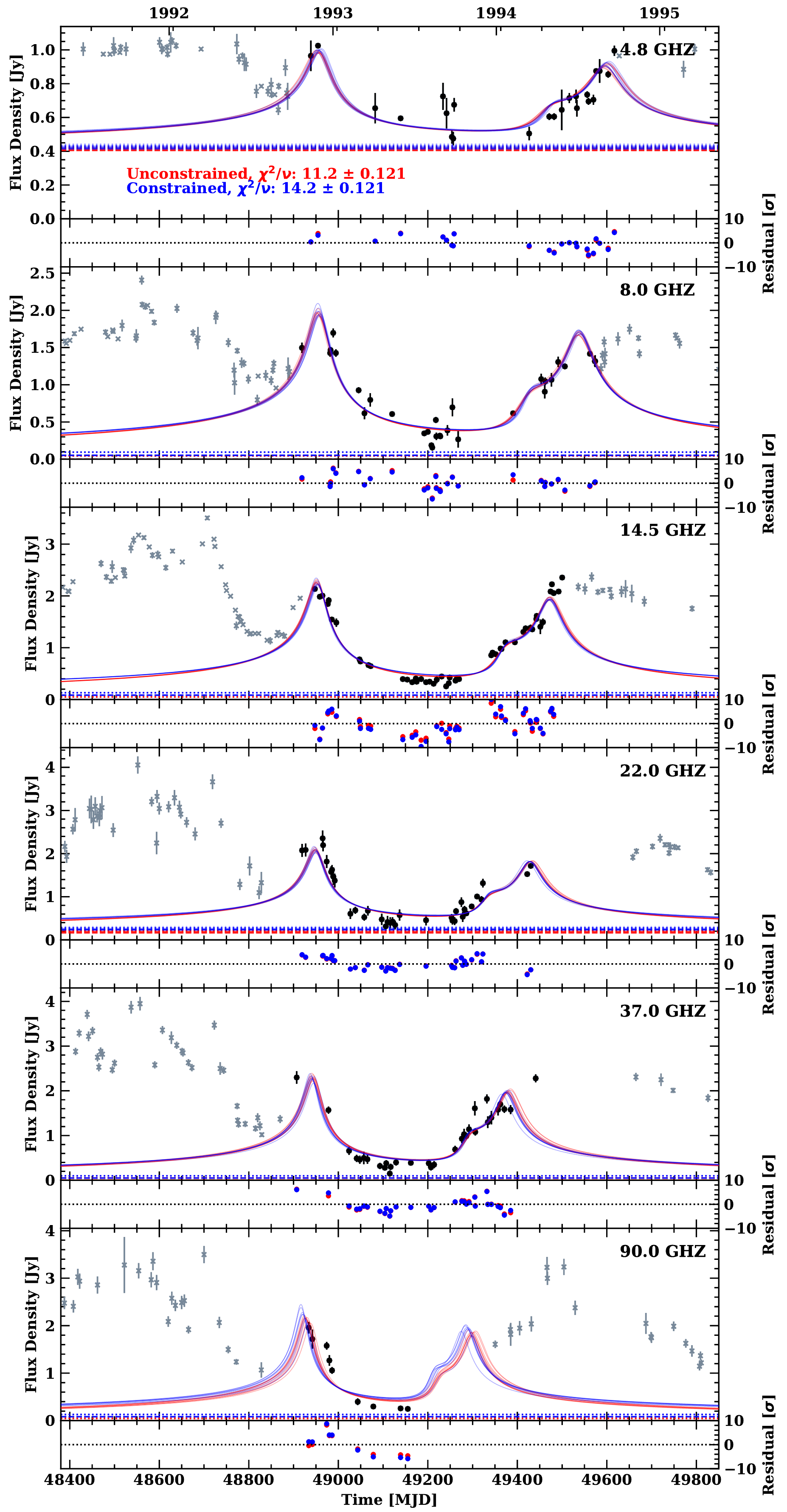}
\caption{Same as fig.~\ref{plt:sav1} but for SAV2.}
\label{plt:sav2}
\end{figure}

\begin{figure}[h!]
\includegraphics[width=0.49\textwidth]{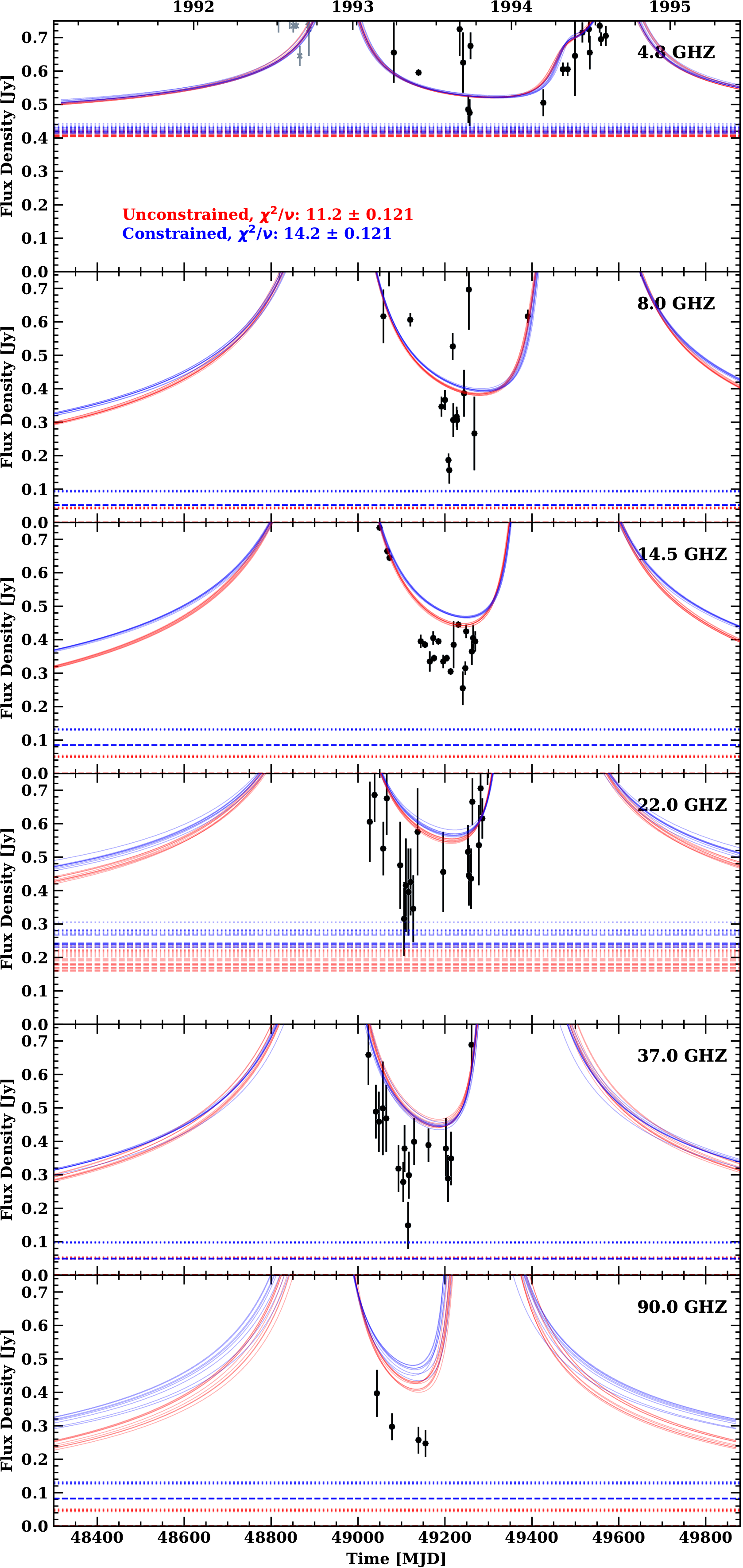}
\caption{Close-up of fig.~\ref{plt:sav2} (SAV2) minimum.}
\label{plt:zoom}
\end{figure}

\begin{figure}
\includegraphics[width=0.5\textwidth]{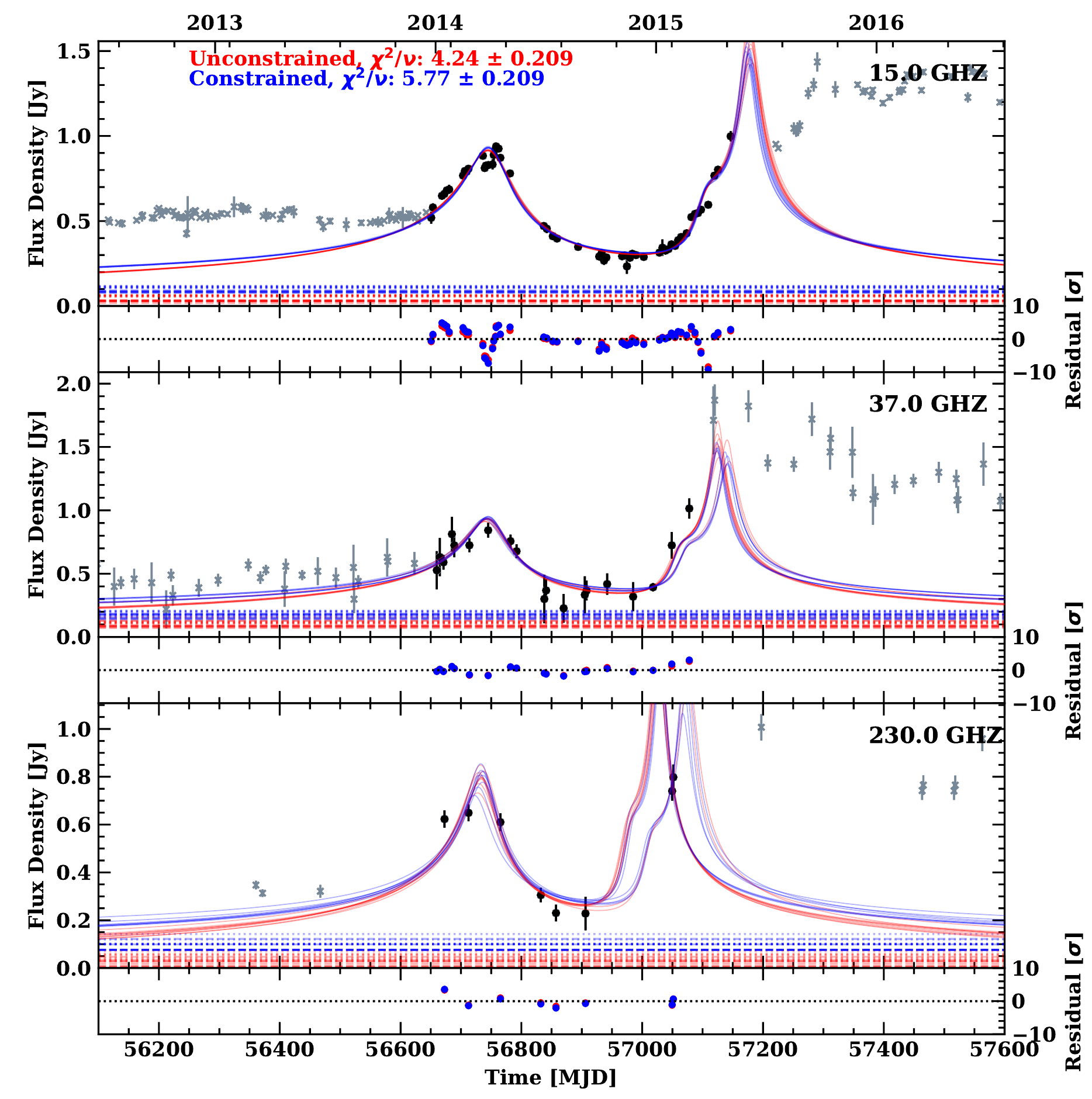}
\caption{Same as fig.~\ref{plt:sav1} but for SAV5. Unlike fig.~\ref{plt:sav45}, SAV5 is treated as a single lensing event: the SAV4 component is not included during fitting.}
\label{plt:sav5}
\end{figure}

\begin{figure}
\includegraphics[width=0.5\textwidth]{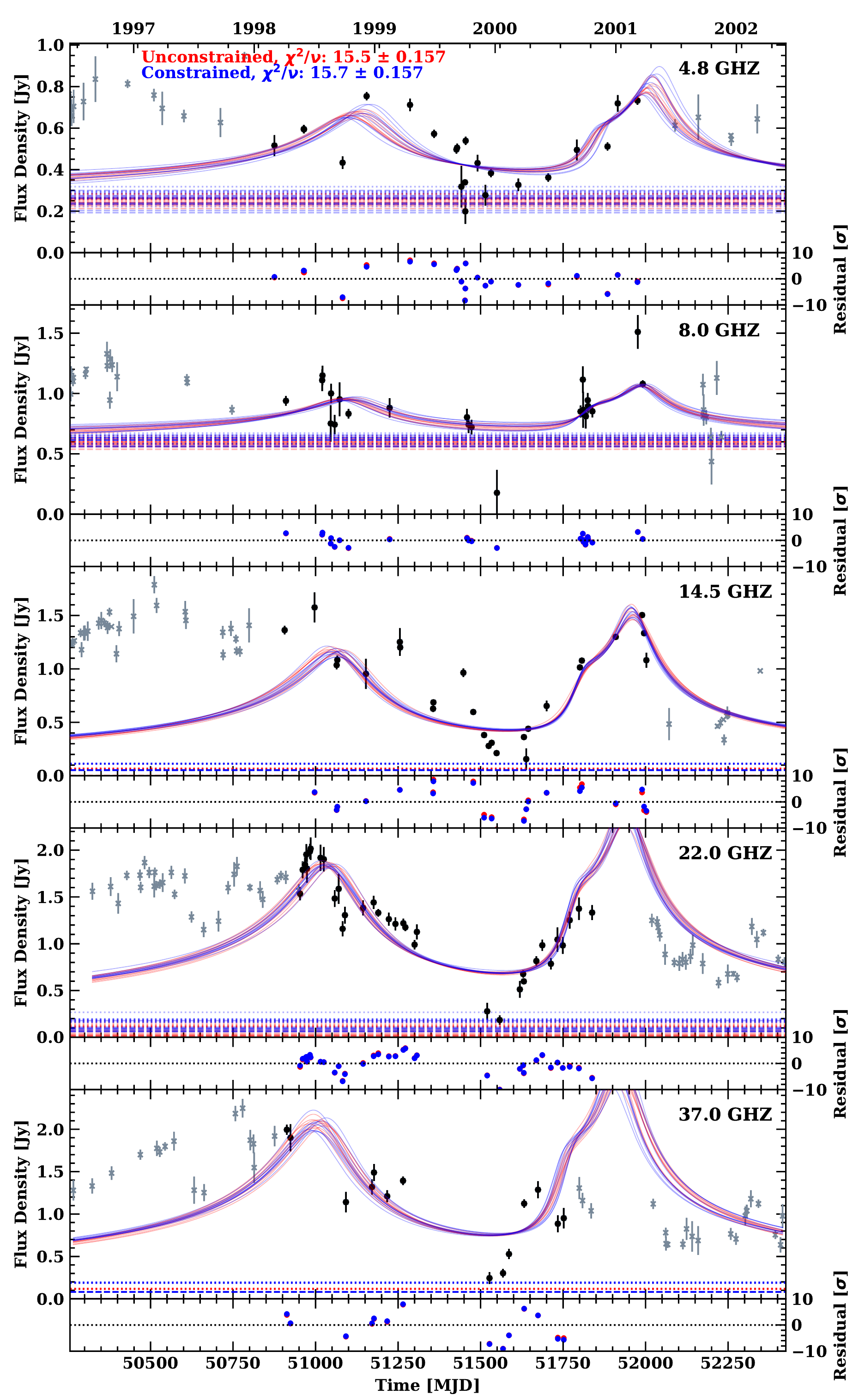}
\caption{Same as fig.~\ref{plt:sav1} but for SAV3.}
\label{plt:sav3}
\end{figure}

\section{Results}\label{sec:res}
If the SAVs found in PKS~1413+135 are indeed millilensing events, we expect them all to be generated by the same stationary and unvarying lens model. To explore that possibility, we attempt to fit all SAVs at once using our binary lens model. {For all fits we use a uniform prior on the fitting parameters that covers the plausible values and excludes degenerate solutions.}
Our only prior constraints are that $k + \gamma \lesssim 0.3$ so the magnification curves are not dominated by a constant external mass distribution. We have verified that a constant external mass distribution with a single point lens is not sufficient to fit SAVs, and Paper~1 showed an elliptical lens distribution is also insufficient.
% All fits include the frequency dependent power law offsets $\eta_0, \eta_E$ for $t_0$ and $t_E$; without this simple SAV frequency dependence, chi-squared values are significantly worse.

We first fitted SAVs 1,2,4 and 5 at all observed frequencies simultaneously. Each SAV, with all relevant frequencies included, is weighted to contribute equally to the likelihood function. The known unlensed components for each frequency are subtracted from the lightcurves before fitting, as described in \textsection2.5. This fit has a total of 29 free parameters: the 5 intrinsic lens parameters, and each individual SAV's source trajectory parameters ($4 \times 6$). Each of these parameters and their priors are detailed in Table~\ref{tab:prior}. The fitting also determines the linear flux density scaling and offset for each frequency and SAV -- these give the core lensed and unlensed flux density.
{Figs.~\ref{plt:sav45}-\ref{plt:sav5}} respectively show the results of the fitting on SAVs for all relevant frequencies. 
{Each colored trace represents a single model realization using an individual parameter sample from the posterior, fig.~\ref{plt:corn}}. 

To prevent overfitting the data, we run two separate fitting procedures. The first, shown in red, assumes the core unlensed flux density can take any value $\geq 0$ (unconstrained). Here the fitting is free to choose that the entire blazar core is being lensed (unlensed flux density $= 0$). The second, shown in blue, assumes the core unlensed flux density must be $\geq 1/3$ of the minimum SAV flux density (constrained). This ensures that the total core flux density is always greater than the unlensed components. {This is usually the case in blazars at 15 GHz. and can be seen to be the case in PKS 1413+135 from 1995-2012, when we have VLBI observations -- see Fig. \ref{plt:components}(a). We discuss this further in \S6.1.}
Each SAV fit result figure displays the total core unlensed flux density level as a dashed line below the observations. Dotted lines show the total core demagnified flux density, i.e. the core flux density that would be observed if there were no magnification. {Again, each line represents a single model realization using an individual parameter sample from the posterior.} 

{For each SAV, only black observations are included in the fitting. The black points were selected by eye based on symmetry and similarity to the low intrinsic noise SAVs 4+5. 
We fitted a handful of different SAV definitions, finding that small changes to the black observation set do not alter fit quality and parameter posteriors significantly.
However, for trustworthy constraints on the lensing system parameters a more thorough investigation is needed, including potentially large changes to the SAV definitions, beyond the scope of this exploratory analysis.}

Table~\ref{tab:val} shows the total fit result for the constrained (blue) case. A corner plot of the posterior over the intrinsic lens parameters is shown in Fig.~\ref{plt:corn}, and we show the source paths behind the lens caustics for each SAV in Fig.~\ref{plt:caustic}.
We find that source components crossing a lens with low binary mass ratio $q \sim 0.006$ and high binary separation $s \sim 1.2\theta_E$ best reproduce SAVs 1,2,4 and 5.

\paragraph{SAV4+5} Fig.~\ref{plt:sav45} shows the fit results for SAVs 4 and 5, explored in Paper~1 using ray-tracing. In this figure, SAVs 4 and 5 are treated as the sum of two lensing events: {their total lensed and unlensed flux density is summed during fitting, so they share a linear flux density offset. We made this simplification, following Paper 1, because the intrinsic (non-SAV) flux density variation is low during this time period and can be mostly removed as a linear trend from the light curve.} 

{Note that the unconstrained (red) model has a smaller unlensed constant flux density than the constrained (blue) model. If unconstrained, it is easier to select lensing parameters that assume more of the flux density is lensed since these require lower magnifications and/or can dip to low flux density more easily. On the other hand, the constrained blue fits require there to be atleast some unlensed flux density, raising the minimum flux density and making more extreme lensing parameters necessary to capture dips.

{Total $\chi^2$ per degree of freedom values for all frequencies are given in the top left of the figure for both models. Since our simplified binary lensing model is likely misspecified, the intrinsic variability of the source is not included in the measurement errors, and we are fitting non-linear models with constraints \citep{andrae2010dos}, these $\chi^2 / \nu$ values should not be taken too seriously. They serve only to compare fit quality between SAVs and models.}

For SAV4+5, the fit quality is visually comparable between frequencies, and SAV5 in particular shows a significant chromatic dependence ($\eta_E << 0$). Higher frequency components have shorter crossing times $t_E$.}

\paragraph{SAV4}
Fig.~\ref{plt:symmetry} shows a close-up of SAV4, to highlight its remarkable symmetry.

\paragraph{SAV1}
{In fig.~\ref{plt:sav1} we show the fit results for SAV1. Looking at the residuals for SAV1, there is variation in fit quality as a function of frequency. Although this can mostly be attributed to small errorbars that don't consider intrinsic variability, the 14.5 GHz clearly misses a peaked feature. This is likely because our simplified lensing model is misspecified: we do not consider finite source size variation as a function of frequency, common in blazars \citep{1979ApJ...232...34B}. Higher frequencies have smaller source sizes and thus more peaked magnifications.

SAV1 shows a similar chromatic dependence ($\eta_E << 0, \eta_0 \approx 0$) to SAV5, although less strong.}

\paragraph{SAV2}
Fig.~\ref{plt:sav2} gives the fit results for SAV2. SAV2 is the most extreme event and covers the widest range of frequencies. It shows the strongest chromatic dependence, in the same form as SAVs 1 and 5 ($\eta_E << 0, \eta_0 \approx 0$). The dependence is strong enough to be visually obvious.

SAV2 dips so low in flux density at higher frequencies that the unlensed and demagnified core fluxes are very low (0.1-0.2 Jy) during the event. It shows potentially significant deviation in some frequencies at minimum; 
fig.~\ref{plt:zoom} shows a close-up of the event minimum at all frequencies. We return to this point in \textsection \ref{sec:difficulties}.

\paragraph{SAV5}
The SAV5 fit without summing the SAV4 component is shown in fig.~\ref{plt:sav5}. Not including SAV4 improves the fit quality, since half a degree of freedom is gained in choosing the unlensed flux density independent of SAV4. The chromatic dependence can be seen by eye.\\

The fit reproduces SAV features {consistently} across frequencies, given the simple lensing model with point source approximation. The source component angles $\alpha$ agree to within $5^{\circ}$. This is consistent with jet components moving along a fixed direction behind a stationary lens. We discuss this in \textsection \ref{sec:jetmodels}.

We are able to achieve a significantly improved model evidence and lower chi-squared values when chromatic time offsets $\eta_0$ and $\eta_E$ are included for all SAVs.
This chromaticity is very clear in the case of SAV2, Figs.~\ref{plt:sav2} \& \ref{plt:zoom}, where lower frequency components have a significantly longer crossing time ($t_E(\nu) \sim \nu^{-0.19}$). Indeed, in all the simultaneously fitted SAVs (1,2,4,5) we find that lower frequency components have longer crossing times and also occur slightly later in time. 

The observed chromatic dependence can be reproduced if the jet flow passing behind a fixed lens is faster and/or further ahead at higher frequency. Transverse velocity variations have been found in both observations and simulations of astrophysical jets (e.g., \citealp{tchekhovskoy_simulations_2008,mertens_detection_2016, mertens_kinematics_2016}). The simplest model suggests a spine-sheath  jet structure, with a faster-moving, more energetic, central spine surrounded by a slower moving sheath (e.g., \citealp{Ghisellini2005}). This kind of structure can also temporarily result from fast shocks or disturbances moving through a slower underlying flow \citep{Blandford1979,ghisellini_rapid_2008}. A finite source size model where the source size varies with frequency should be used to investigate these potential structures in detail.
%High frequency part needs to come first.

The quality of fit and parameter posteriors are of course somewhat dependent on the definition of SAVs, Because of this ambiguity in SAV definition, our uncertainty in the final lens parameters is larger than suggested by their posteriors in Table~\ref{tab:val}.
We do not expect perfect fits across all frequencies since the source clearly has a large amount of variability on top of the lensing that our model does not account for. Indeed, blazars are by nature variable on all timescales and these variations are not captured by the displayed flux density measurement uncertainties. Furthermore, magnification through millilensing will increase absolute variations in the original source; Fig.~\ref{plt:J0920+4441} shows an example. 
% Our model also does not consider finite source size effects that may blur magnification patterns for lower frequencies. 

\subsection{Milli-lens Fitting of the Anomalous SAV3}

We fit SAV3 separately, fitting it jointly with SAV1,4,5 in order to constrain its lensing parameters.
Both the constrained and unconstrained model fits are unsatisfactory, especially at 22 GHz and 37 GHz.  
SAV3 is the most sparsely sampled event, shows the highest variability of all the events and is the only event showing a positive chromatic component $\eta_E$. Furthermore, the large flux density external to the core during SAV3, fig.~\ref{plt:constraints}, and its low demagnified core component, fig.~\ref{plt:sav3}, make it the only event where the core is less bright than the external components, \S\ref{sec:difficulties}.
These unsatisfactory fits, SAV3's asymmetry fig.~\ref{plt:components}, and the required core dimness suggest that SAV3 is dominated by intrinsic variability rather than gravitational lensing.
%We exclude SAV3 from our final model for the lens...

\section{Three Jet Models}\label{sec:jetmodels}
\begin{figure}[h!]
\includegraphics[width=0.5\textwidth]{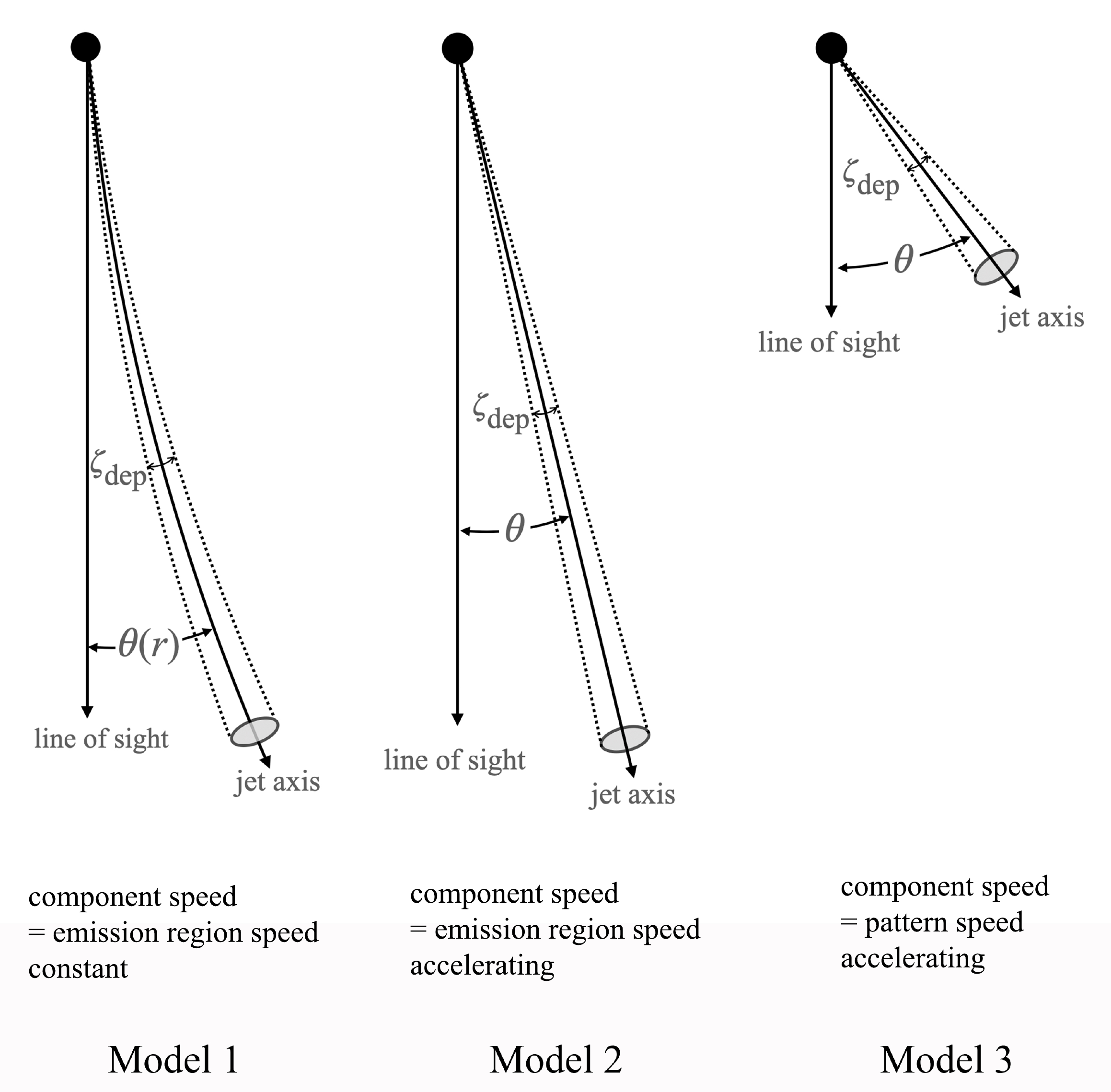}
\caption{Schematic of Models 1, 2 \& 3. In Models 1 \& 2 the apparent speeds of jet components are related to the speed of the emission regions and hence to $\theta$. In  Model 3 the apparent speeds of the jet components  are pattern speeds  unrelated to $\theta$. Thus on Model 3 $\theta$ can be greater than in Models 1 \& 2.} 
\label{plt:models}
\end{figure}

\begin{figure*}[ht!]
\centering
\includegraphics[width=0.8\textwidth]{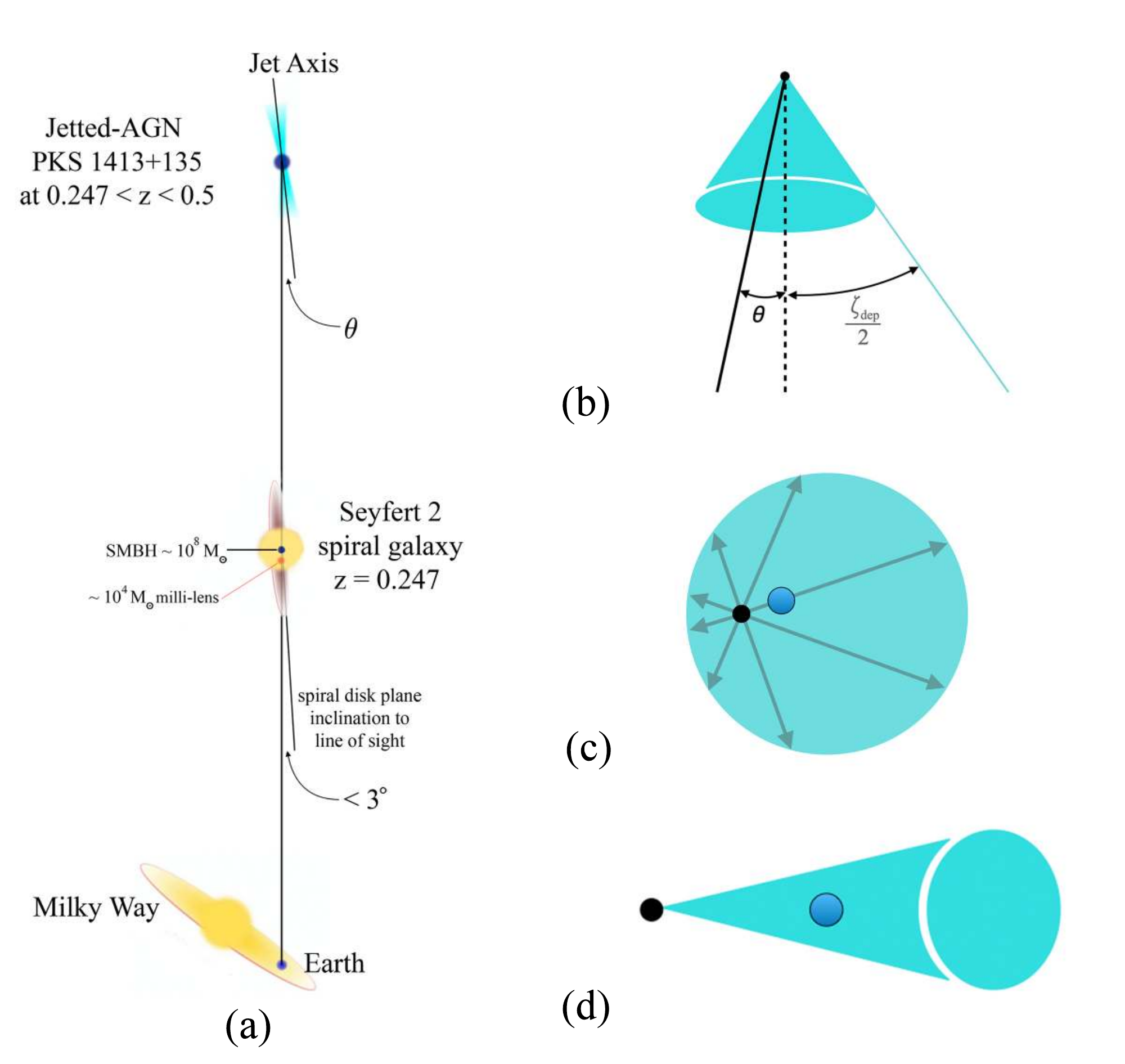}
\caption{(a) The orientation and suggested configuration in Paper 3 of the blazar PKS\,1413+135  and the Seyfert 2 spiral galaxy. The angle $\theta$ indicates the angle between the jet axis and the line of sight. (b) The relationship between $\theta$ and the semi-angle of the jet cone $\zeta_{\rm dep}/2$. (c) the situation if $\theta<\zeta_{\rm app}/2$, i.e. the line of sight lies within the jet cone. (d) the situation if $\theta>\zeta_{\rm app}/2$, i.e. the line of sight lies outside he jet cone. In (b), (c) \& (d) the black disk marks the position of the central engine. In (c) \& (d) the dark blue disc has a radius equal to the Einstein radius of the millilens, and hence represents the approximate apparent size of the lens. In (c) only a small fraction of trajectories from  the central engine intersect the lens. In (d) a significant fraction of trajectories from  the central engine intersect the lens.}
\label{plt:Cartoonsav}
\end{figure*}

\begin{figure*}[ht!]
\centering
\includegraphics[width=1.0\textwidth]{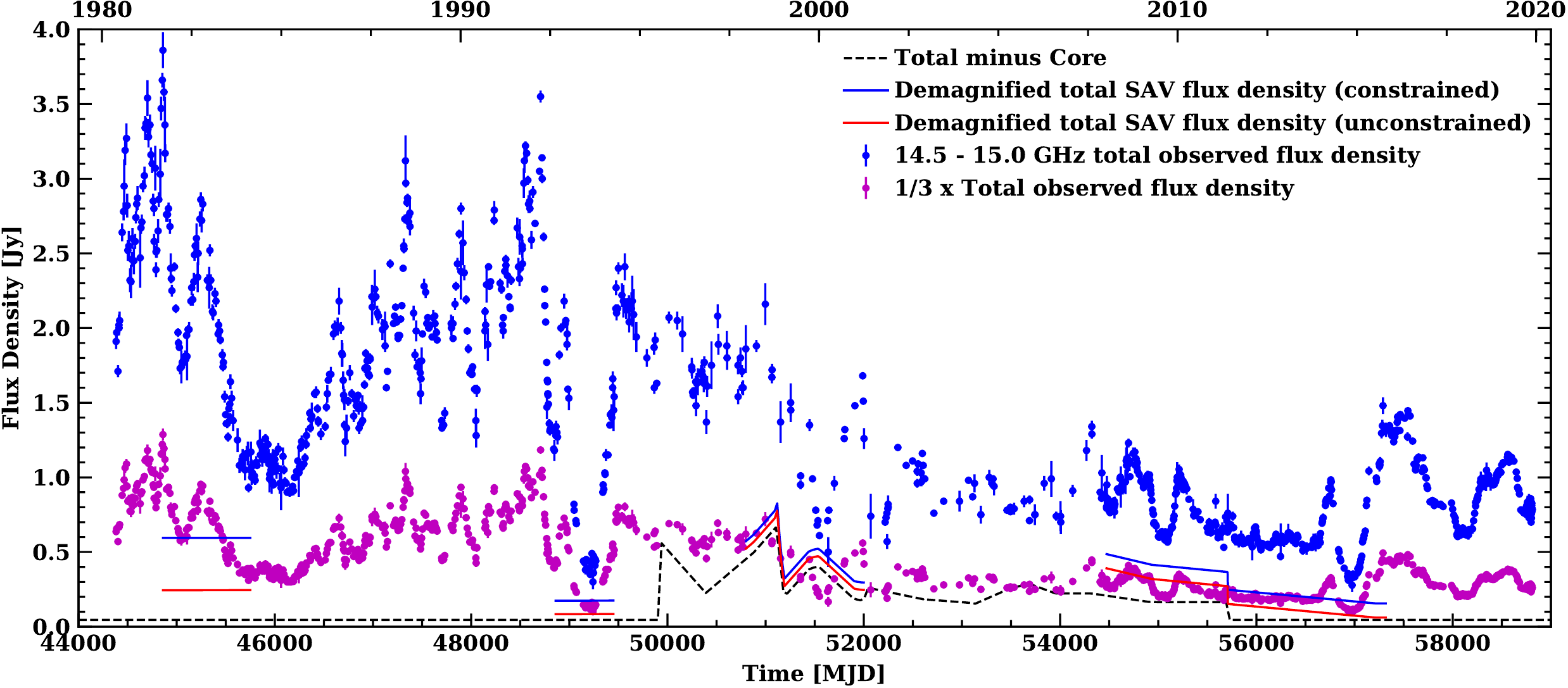}
\caption{The combined UMRAO 14.5 GHz and OVRO 15 GHz flux densities from 1980 to 2020 (blue circles). During the period of overlap from MJD 54473 to MJD 55759 the OVRO data are used. During SAV1, SAV2, SAV4 \& SAV5 the de-magnified flux densities for both the constrained and unconstrained fits are shown by the blue and red lines respectively. 
%The green points show the observed minus the model values. 
%The open black squares show the total flux densities measured by MOJAVE. 
The dashed black line shows the total minus the core flux density, indicating the flux density level that is certainly unlensed. 
%The open black circles with crosses in them -- which have a mean of 44 mJy -- show the total flux density minus the core minus D8 minus D7  (see Fig. \ref{plt:collimation}). 
The purple crosses show the 14.5 GHz UMRAO and 15 GHz OVRO flux densities reduced by a factor 3. This is a plausible guess at what the PKS 1413+135 lightcurve might look like outside of the observed 
SAVs in the total absence of milli-lensing  (see text). }
\label{plt:demagnified}
\end{figure*}

The gravitational lensing interpretation of SAV is clearly only viable if it is possible to construct  models of the jet in PKS 1413+135 that are consistent with all the observations and the lensing model derived from  SAVs.

In this section we discuss three different models for interpreting the observations, one of which, as shown in \S \ref{sec:model1}, is ruled out if SAV is caused by gravitational millilensing. The models are shown in Fig. \ref{plt:models}. In all three models we assume that we are dealing with a circularly symmetric conical jet, and that the milli-lensing occurs within the core, as discussed in \S \ref{sec:origin}. 
We denote the apparent (i.e. the observed) jet opening angle by $\zeta_{\rm app}$. This is related to the deprojected angles depicted in Fig. \ref{plt:models} by  $\zeta_{\rm app}=\zeta_{\rm dep}/{\rm sin}\theta$. In Model 1 $\theta$ depends on $r$, the angular distance from the core. In Models 2 \& 3 $\theta$ is constant. In Models 1 \& 2 the apparent speeds of components in the jet are assumed to be due to motion of the emission regions, and hence related to $\theta$. In Model 3 the apparent speeds of components in the jet are assumed to be pattern speeds, and hence unrelated to $\theta$ \citep{1985ApJ...295..358L,2007ApJ...658..232C}.

\subsection{Model 1}\label{sec:model1}
In Model 1 we assume that the apparent speeds observed in the VLBI observations are speeds of the emitting regions that are moving with the same speed as the bulk  flow speed of the material moving along the jet.

The results of \citet{2019ApJ...874...43L} show that between $r=0.32$ mas (D8) and  $r=7.15$ mas (D3) the apparent speed of the jet components increase with $r$ from $\beta = 0.144\pm 0.034$ to $\beta = 1.72\pm 0.011$, assuming $z=0.247$, where $\beta=v/c$. These numbers are increased to $\beta = 0.228\pm 0.054$ to $\beta = 7.72\pm 0.017$ for $z=0.5$.

In Paper 3 we showed that the large scale structure of PKS 1413+135 is a curved jet, and we showed above that between D8 and D3 the jet curves through angle $\delta \xi{\rm (D3)} = 4.4^\circ\pm1.3^\circ$. Thus the simplest explanation for the increasing apparent speed of the jet is that the speed of the material moving down the jet is constant, but the jet is is curving away from the line of sight between D8 and D3, which is the interpretation we adopted in Paper 3.
We now examine this model in more detail. 

On this model in Paper 3 we showed that the angle of the jet to the line of sight at the position of D8 is $\theta \lesssim 1.4^\circ$ (see Table 5 of Paper 3). Thus, given the
jet width of $11.1^\circ \pm 0.5^\circ$ determined by  \citet{2017MNRAS.468.4992P}  (see \S \ref{sec:opening}), at the position of component D8 the line of sight lies within the cone, so material from the core streams out in all directions on this model, as shown  in Fig. \ref{plt:Cartoonsav} (c). On this model, in order to produce SAV, the emission regions ejected from the core must be much smaller than the width of the jet and  most of the material streaming through the core would be unlensed, as can be seen in Fig. \ref{plt:Cartoonsav} (c).   It is unlikely that this would  produce recurring SAVs that dominate the light curve. 
For these reasons we turn to Model 2.

\subsection{Model 2}\label{sec:model2}

In Model 2 we assume that the increase of apparent speed with angular distance from the core is due to acceleration of the observed component \citep{2009ApJ...706.1253H,2015yCat..17980134H}, and that the angle between the jet axis and the line of sight is constant, as shown in Fig. \ref{plt:models}.

In Model 2, as in Model 1, we assume that the apparent speeds observed in the VLBI observations are speeds of the emitting regions but in this case we do not assume that the speed is the same as the bulk flow speed along the jet,  but that the components we are measuring are features that are accelerating, as is observed to be the case in M87 \citep{2018ApJ...855..128W}. Thus in Model 2 we can have $\theta > \zeta_{\rm app}/2$, and trajectories like those between paths 1 and 2 in Fig. \ref{plt:cartoon} can give rise to SAVs, as shown of Fig. \ref{plt:caustic}. This situation, illustrated by \ref{plt:Cartoonsav} (d), provides an entirely plausible interpretation of the observations.

\subsection{Model 3}\label{sec:model3}

As pointed out in Paper 3, there is evidence that components B and C in the counterjet are moving away from the core at speeds of $\beta = 0.7 \pm 0.4$ and $\beta = 0.32\pm  0.19$, respectively, for an asumed redshift of z=0.247. These are marginal detections of motion, but if they are real then the apparent speed cannot be due to motion of the emission regions, since the counterjet  is pointing away from us, so these must be pattern speeds \citep{1985ApJ...295..358L,2007ApJ...658..232C}, and tell us nothing about the orientation of the jet relative to the line of sight.   For this reason we now consider Model 3.

Model 3 is very similar to Model 2, but there is no problem with relativistic speeds in the counterjet, should these prove to be real. On Model 3, therefore, we place no constraints on the angle between the jet axis and the line of sight based on component speeds, but only apply the usual rule for the beamed emission from a blazar jet that the angle between the jet axis and the line of sight is not $\gg 1/\Gamma$, where $\Gamma$ is the Lorentz gamma factor.

\subsubsection{The Apparent Speeds of Components B and C}\label{sec:apparent}
Part of the motivation for Model 3 is the fact that, as discussed in Paper 3, the apparent speeds of components B and C are $\beta_{\rm app} = 0.7 \pm 0.4 \; {\rm and} \; 0.32c \pm 0.19c$, respectively, and therefore possibly relativistic. If they are indeed relativistic this would prove that these are pattern speeds because a relativistic apparent speed in a counterjet can only happen if the jet axis is nearly orthogonal to the line of sight, which we showed in Paper 3 is not the case in PKS 1413+135. Given the sizes and low surface brightness of components B and C shown in Fig. \ref{plt:structure} it will take many decades to measure the speeds, or upper limits on these, to the accuracy required.  We have therefore re-examined the 23 epoch MOJAVE data with the following results.

In component B the apparent motion is entirely driven by epochs 1999-11-06 and 1999-12-27, which are 0.5 mas away from all the rest, which casts doubt on their reliability.  Component C is not visible at either of those two epochs, which also casts doubt on the quality of the images. Estimating errors on individual epoch positions is virtually impossible, due to non-linear effects of antenna dropouts, uv coverage, and self-calibration.  As a result we find that component B is consistent with zero apparent motion from 1999 to 2011.5.

In feature C, the acceleration fit (and speed) is being driven by a single epoch (1999-01-09), which is 0.4 mas from all the others.  This is the weakest feature in the source, and thus has the largest positional error. It looks as if  the data points are moving back and forth around a single position, consistent with zero motion.  In general we view any MOJAVE speeds with significance $<2\sigma$  with suspicion, since the errors  on each individual point are not well known.  Thus the situation for component B and C in the counterjet is very different to the situation for components D8, D7, D6, D4 \& D3 in the jet, which, as can be seen in Fig. \ref{plt:collimation}, can be determined with  high precision from the MOJAVE images.

While this does not disqualify Model 3, it does remove any evidence for this model based on component speeds.

\section{Difficulties with the lensing hypothesis}\label{sec:difficulties}
In this section we discuss two potential problems with the lensing analysis that we have carried out when compared with the observed light curves of PKS 1413+135 shown in Fig. \ref{plt:lightcurves}.

\subsection{The Levels of the Unlensed and Demagnified Components of PKS 1413+135}\label {levels}

The effect of demagnifying the putative lensed components during SAV1, 2, 4 \& 5 is shown in Fig. \ref{plt:demagnified}. 
The abrupt jumps in flux density level seen in Fig. \ref{plt:demagnified} (from the blue dots to the blue or red lines) at the times of the putative lensing events is highly implausible, since it would require the flux densities of unlensed components in the core to drop simultaneously with the transit behind the lens of the lensed components in the core.

The demagnified flux density level during SAVs is much lower than the average flux density, whereas gravitational lensing amplifies the signal of the lensed component and does not affect the unlensed components so that outside of the window when the lensing is occurring the signal is expected in general to be lower than the signal during the lensing event.

\begin{figure}[h!]
\includegraphics[width=0.5\textwidth]{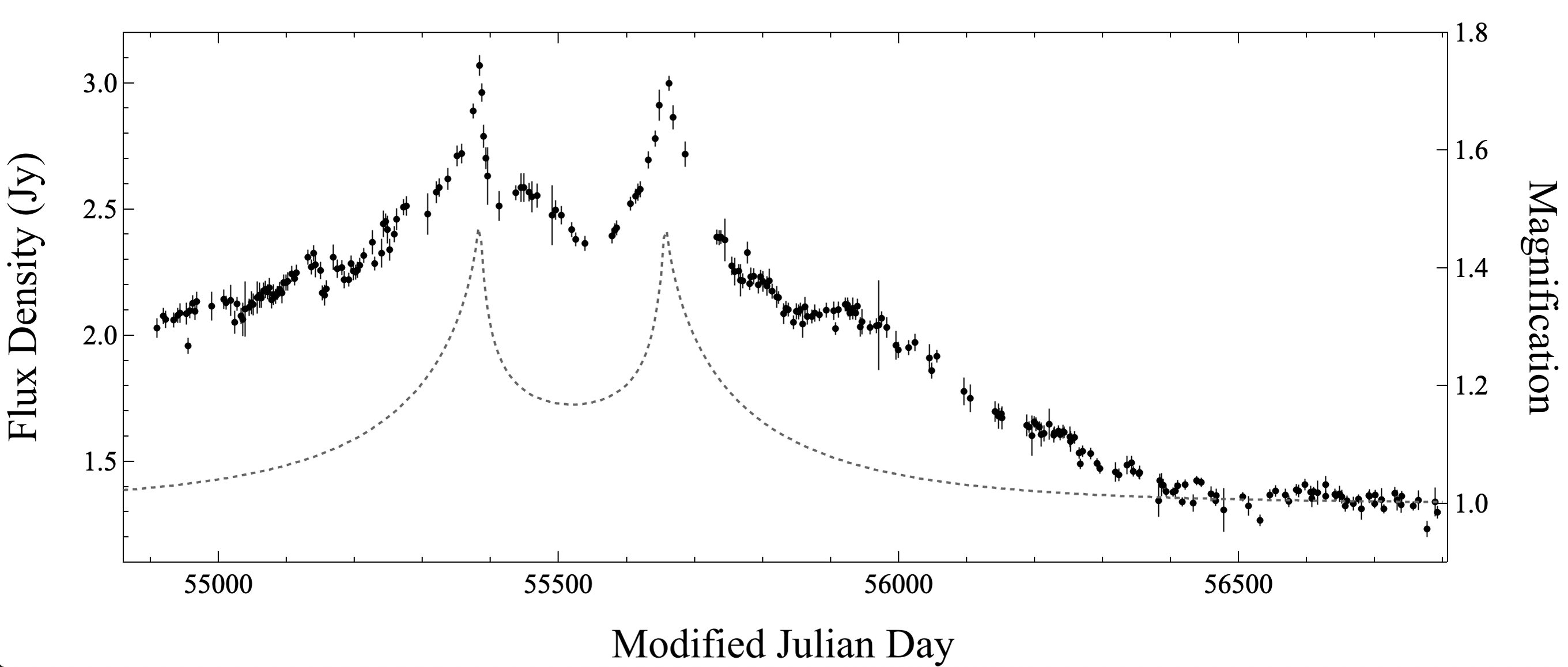}
\caption{An example of a blazar 15 GHz light curve (J0920+4441) to which a single unresolved component has been added, illustrating the increase in flux density. In this case the component fades away about a year after the transit behind the lens. Note that the absolute variability is magnified by lensing. The magnification is shown by the gray dashed curve and right-hand scale.}
\label{plt:J0920+4441}
\end{figure}

An example is shown in Fig. \ref{plt:J0920+4441}, where we have added to the OVRO lightcurve of the blazar J0920+4441 an unresolved component that transits behind a putative lens similar to the lens in the case of PKS 1413+135. In this case the component fades after transiting behind the lens. This illustrates the point that in general gravitational lensing boosts the lightcurve above the surrounding levels.

The black dashed line in Fig. \ref{plt:demagnified} shows the combined flux densities of the components outside of the core, which are unlensed. 
%So on the hypothesis that apart from the flux density level of these points all the remaining flux density, i.e. the difference between the blue or red lines and the black points, is lensed, we can see that we are overfitting the data during the lensing events because the blue light curves during SAVs are much less variable than the black points are during the period 1995 - 2008 when there are no  identified SAVs.  The difference between the model fits and the observed data are shown by the green points during SAVs. These differences give a more realistic estimate of the variability of the unlensed components during SAVs. 
In our unconstrained model fit (red) we assumed that all of the flux density of the core could be lensed during SAVs.  However, were this true, then the demagnified flux density of the core during these SAVs would be much less than the flux density of the principal two components, D7 \& D8, in the jet. This would be highly unusual, since at 15 GHz the compact core usually dominates the jet in blazars with high variability Doppler factors.

While we cannot determine what fraction of the flux density of the core is unlensed, it is instructive to consider the case where, for example, one-third of the total flux density is unlensed.  In Fig. \ref{plt:demagnified}  the 14.5 GHz UMRAO and 15 GHz flux densities of PKS 1413+135 are shown reduced by a factor 3 by the purple circles.  Regions where we have overfitted SAV features, assuming that the unlensed components account for one-third of the total flux density, are those where the blue or red lines of the de-magnified flux densities fall below the purple circles.

The demagnified light curves for our constrained model fit are shown by the blue lines in Fig. \ref{plt:demagnified}. This model was constrained to ensure atleast 1/3 of the core flux density is not magnified by the lens. For SAVs 1,2,4,5 the constrained model gives more reasonable demagnified core flux densities that are significantly brighter than the unlensed components outside the core, the tradeoff being slightly higher $\chi^2$ values for the fits (fig.~\ref{plt:sav1} - \ref{plt:sav5}). SAV3 is once again the outlier; its core flux density does not exceed the components outside the core even under the constrained model.

In Paper 3 the variability Doppler factors  were derived on the assumption that the lightcurve of  PKS 1413+135 outside of SAVs is not significantly affected by lensing, but as we have shown in \S \ref{sec:Doppler}, based on the variability of component D8, which we know is unlensed, and which dominates the unlensed curve given by the gray crosses in Fig. \ref{plt:constraints}, these variability Doppler factors are in good agreement.  This is important since the variability Doppler factor was the major argument in Paper 3 in the determination of the orientation of the jet axis of PKS 1413+135 relative to the line of sight. So that conclusion is not changed if the light curves of PKS 1413+135 are dominated by lensed components.

\subsection{The low points in SAV2}\label{sec:low}
The lowest observed flux densities in SAV2 occur near the center of the SAV and these are significantly lower than the model at 14.5, 37.0 and 90.0 GHz. We have not been able to fit these with our lensing model. The only way to reconcile them with the lensing model is to assume that the flux density of the lensed component decreased by $\sim$25\% at frequencies from 14.5 GHz - 90 GHz, and then increased again.  This is by no means impossible or even unlikely, especially considering the large ($> 25\%$) variability at SAV2 minimum in the 37GHz and 8GHz light curves, but it does require a decrease and increase in the flux density that mimics the lensing effect, which may cast some doubt on the lensing hypothesis in the case of  SAV2.
%Finite source size effects.

\section{Candidate Millilenses}\label{sec:lenses}
In Paper 3 we showed that the jetted-AGN PKS 1413+135 is almost certainly located behind the spiral galaxy. But we could not definitively rule out the possibility that it is located in the spiral galaxy and powered by the $\sim 10^8 \; M_\odot$ supermassive black hole (SMBH) responsible for its Seyfert 2 characteristics of the spiral.  In this section we discuss two possibilities: (i) that the jetted-AGN is a background source and that the putative millilens is the GMC for which the evidence has been reported by \citet{2002AJ....124.2401P}; and (ii) that the jetted-AGN is located in the spiral galaxy.

\subsection{A GMC in the Spiral as a Millilens}\label{sec:GMC}

\citet{1996AJ....111.1839P,2002AJ....124.2401P} have presented evidence for a giant molecular cloud (GMC) in the foreground spiral galaxy  along the line of sight to the blazar PKS 1413+135. The largest GMCs have virial masses of $10^7 M_\odot$ and typical sizes of 30 pc \citep{2010ARA&A..48..547F} and so have projected densities of $ \approx 10^4 M_\odot \; {\rm pc}^{-2}$, i.e. the GMC in the spiral galaxy discussed by \citet{1996AJ....111.1839P,2002AJ....124.2401P} could well have the projected density required for the lens that we postulate is responsible for SAV in PKS 1413+135. This would not be surprising given that millilensing by GMCs has been found to be more common than other structures in intervening spiral galaxies (e.g., \citealp{Sitarek2016}).

\subsection{A Dwarf Galaxy and massive Black Hole as a Millilens}\label{sec:dwarf}

The possibility that the jetted-AGN is located in the spiral galaxy was discussed in Paper 1 \S 6.1, where we showed that our estimated millilens optical depth of $ 10^{-6} \rightarrow 6 \times 10^{-4}$ would require a population of millilenses with $\Omega_l/\Omega_m = 10^{-3} \rightarrow 10^{-1}$.
In Paper 3 we discussed the evidence favoring the hypothesis that the jetted-AGN PKS 1413+135 is located behind the spiral galaxy and we concluded that, while this was almost certainly the case, we could not definitively rule out the possibility that the jetted-AGN is located in the spiral Seyfert 2 galaxy and powered by its $\sim 10^8\; M_\odot$ SMBH.

The evidence favoring the hypothesis that the jetted-AGN is located in the spiral galaxy is the following.
\vskip 6pt
\noindent
1. As shown in Paper 3, the probability of the alignment, to within $13\pm 4$ mas \citep{2002AJ....124.2401P}, of the jetted AGN and the centroid of the infrared isophotes of the spiral is $1.0 \times 10^{-4}$. \citet{2002AJ....124.2401P} showed that the probability of the alignment of a GMC, with projected dimensions of $\sim 1 \times 15$ kpc for the dust lane containing the GMC, is $\sim 2 \times 10^{-4}$.  Thus we have a probability of $\sim 2 \times 10^{-8}$ to contend with under the hypothesis of \S \ref{sec:GMC}. Although multiplying probabilities is an unreliable procedure, it is undeniable that in the case of the hypothesis of \S \ref{sec:GMC}
we have two unlikely alignments that require explanation, whereas under the hypothesis that the jetted-AGN is located in the spiral there is only one, namely the alignment of the jetted-AGN with the lens, requiring a single {\it a posteriori\/} low probability. 
\vskip 6pt
\noindent
2. If the jetted-AGN lies behind the spiral, then it is curious that neither the spiral galaxy itself produces multiple images on the $\sim 1$ arcsecond scale, nor does the $\sim 10^8\; M_\odot$ SMBH in the spiral produce multiple images on the scale of $\sim$ 10 millarcseconds.  In Paper 3 we showed that these two facts can be explained through a soft potential for the spiral and possibly greater misalignment of the SMBH, but nevertheless these issues remain a valid concern with the interpretation we favored in Paper 3. 
\vskip 6pt
\noindent
3. As can be seen clearly in Fig. \ref{plt:structure}, the counterjet is a strong radio source. It is very rare for a blazar or a BL Lac object to have a visible counterjet. In Paper 3 we suggested that this is due to interaction between the counterjet and the surrounding medium, such that the emission regions in the counterjet are not moving at relativistic speeds away from us.  If the jetted-AGN lies in the spiral, then the counterjet axis lies within $3^\circ$ of the plane of the galaxy, and this would provide a simple explanation for the strong interaction between the counterjet and the surrounding medium.  

If the jetted-AGN is in the spiral and if SAV is indeed due to millilensing, then since, as shown in Paper 1, the millilens cannot be located in either the spiral galaxy or the Milky Way, we are looking for an intergalactic millilens. The most likely host would therefore be a dwarf galaxy.

\citet{2014ApJ...787L..30R,2020ApJ...888...36R} and \citet{2019ApJ...884...78L} have carried out searches for  massive black holes (MBHs) in dwarf galaxies. They present evidence that 13 of the 39 dwarf galaxies in which they detected compact radio sources  contain MBHs with masses in the range $M_{\rm BH}\sim 10^{4.1} - 10^{5.8}\;M_\odot$. In the majority of the cases the MBHs are offset from the centers of the dwarf galaxies, and their results indicate that MBHs do not always reside in dwarf galaxies. In one instance there is evidence for two MBHs in the same dwarf galaxy.  Their search covered redshifts $z<0.055$.  Given that the spiral galaxy is at $z=0.247$, the dwarf galaxy that we seek could be $3 \times$ further away and an order of magnitude fainter than the systems studied  by the above authors. The detection of such a system along the line of sight to the spiral galaxy would be difficult.

\section{Possible Periodicity and future SAVs}\label{sec:periodicity}

\begin{figure*}
\includegraphics[width=1.0\textwidth]{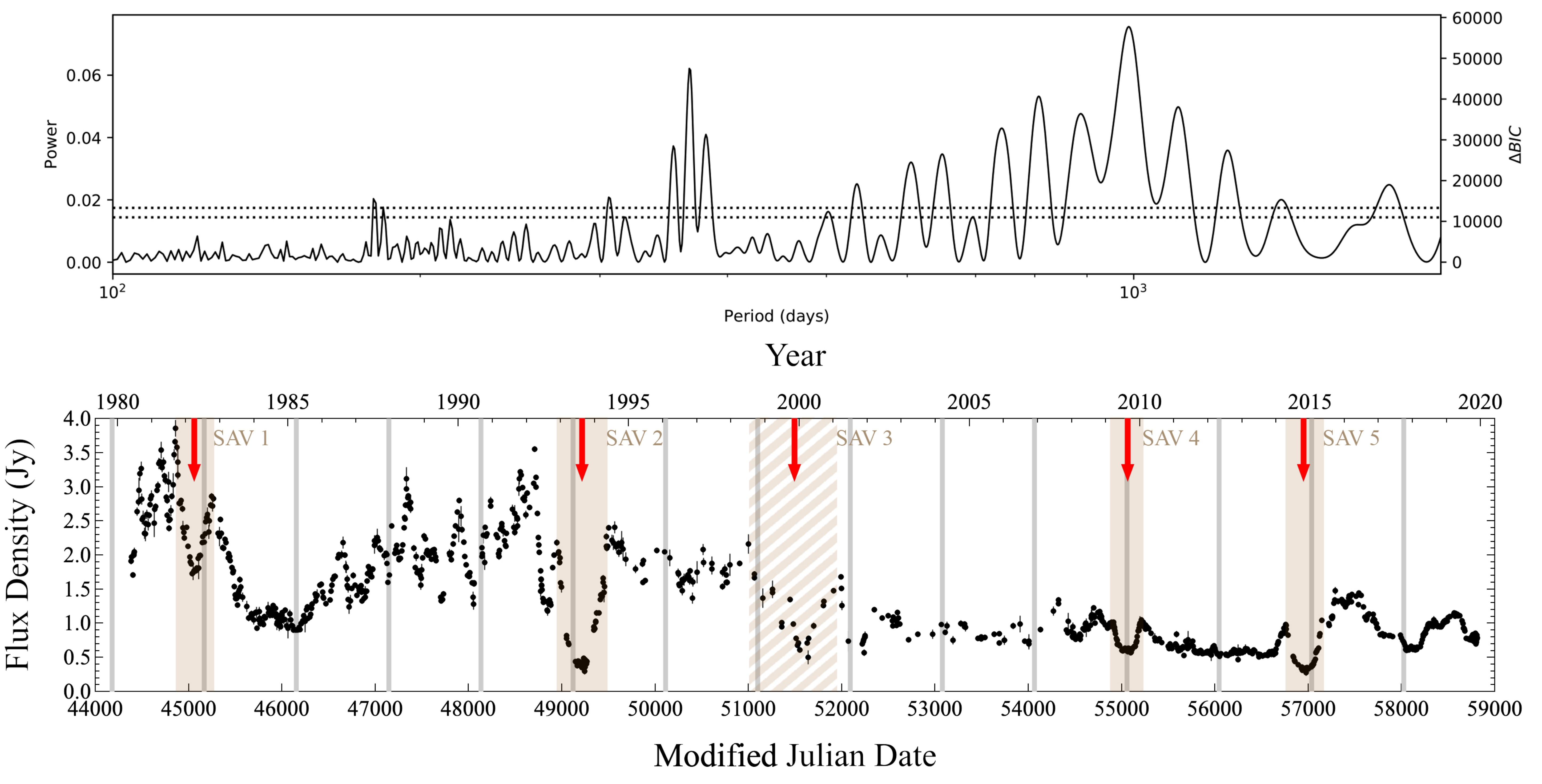}
\caption{Possible periodic behavior in the flux density of PKS 1413+135. Upper plot: Lomb-Scargle analysis of the combined UMRAO 14.5 GHz, OVRO 15 GHz, and AMI 15 GHz data, showing a peak at a period of 989 days. Lower plot: The PKS 1415+135 data used in the Lomb-Scargle analysis, with a comb of spacing 989 days aligned with the SAVs, for which the midpoints between the cusps are indicated by the red arrows.  Note that SAVs in 1982, 1993, 2009 and 2014, corresponding to SAV 1, SAV 2, SAV 4, and SAV 5 fit the comb quite well, whereas that in 2000, corresponding to SAV 3, does not.}
\label{plt:pred}
\end{figure*}

Periodic behavior in AGN can naturally arise from a number of different mechanisms including supermassive black hole binaries, Lense–Thirring precession, the accretion disk and inner jet (as in quasi-periodic oscillations),  and plasma instabilities in the jet. There has been an increasing interest in identifying periodic sources driven by the advent of gravitational wave experiments such as LISA and pulsar timing arrays. In blazars, despite extensive searches, only a handful of sources have shown evidence for periodicity. Notable examples are that of OJ~287 whose binary black hole system shows outbursts with a $\sim12$yr cycle \citep{Valtonen1988} and PG 1553+113 with a tentative $\sim2$yr periodicity at high-energies \citep{PG1553_2015,Raiteri2017}.

Here we consider the possibility of a wobbling jet that would cause SAVs at quasi-periodic intervals. In this scenario, the source would appear as a regular blazar until the jet nutation would align it with the lensing mass in the intervening galaxy causing SAVs. Using the Lomb-Scargle periodogram \citep{Scargle1982} we identify several peaks, the most prominent of which is centered at $\sim989$ days (Fig. \ref{plt:pred}). Assuming $t_0$ to be the minimum of SAV4. our most symmetric SAV, we can plot the expected SAV dates (vertical lines Fig. \ref{plt:pred}). Interestingly, SAV1, SAV2 and SAV5 happened at the dates predicted by this model. The anomalous SAV3 once again does  not fit the pattern of the other four SAVs. There are predicted dates that do not result in visible SAV behavior. This could be the result of imperfect alignment between the background blazar and the lens or that blazar variability from downstream components dominated the emission over the core. It is possible that the apparent periodicity of the SAVs is purely coincidental. But if not then we have predictions for the next three upcoming  SAV wndows: beginning in August 2022, May 2025, and April 2031. If SAV occurs in any of these windows it will not only  conform the periodic behavior in the lightcurve of PKS 1413+135, but it will also confirm that SAV are not simply random  intrinsic variations in brightness, but are caused by some other repetitive behavior that produces a highly distinctive pattern.

\section{Discussion}\label{sec:disc}
%SAV1 (fig.~\ref{plt:sav1}) shows good agreement with the lens model found for SAV4 and 5 

There are two major components of this study -- one is the development of the computer tools for fitting multiple gravitational lensing events at multiple frequencies simultaneously, the other is the implications this has for the millilensing hypothesis in PKS 1413+135.  We discuss these separately below.

In the course of this study we have developed a robust nested sampling methodology to fit multiple gravitational millilensing events found in multiple frequencies. Our results allow us to differentiate between events dominated by intrinsic long-term variability typical of blazars and events dominated by gravitational lensing. Blazar millilensing can be challenging due to the short variability timescales and  light curve sampling that can affect the fitting results. In this work we selected SAV definitions for fitting by inferring the general shape from the less variable SAVs (4+5) and by retaining as much symmetry as possible. 
Within these qualifications, we have shown it is possible to fit the four bona fide SAVs with the same gravitational lens model. 
It is not surprising that the anomalous SAV3 candidate does not fit the lensing model of the four bona fide SAVs because it is clearly dominated by intrinsic variability and not lensing.

Paper 1 placed constraints on the mass of the lens system in this foreground scenario, with a range of mass $\sim 10^2 - 10^5 M_\odot$ weakly dependent on the source distance and strongly dependent on the source angular size.

The brightness temperature  provides an estimate of the angular size of the lensed component, and hence of the lensing mass required to reproduce the observed SAVs. Paper 1 showed that in PKS 1413+135, for the case of an intergalactic lens, these ranged from $10^3-10^4 \; M_{\odot}$ for a brightness temperature of  $10^{12} $K down to $10-100\;M_\odot $ for a brightness temperature of  $10^{14} $K. \citet{2005AJ....130.2473K} studied 250 flat spectrum  sources with multi-epoch VLBI and found that half of them showed components with brightness temperatures exceeding $10^{12}$ K at some epochs, and they report brightness temperatures in the unresolved cores of some blazars that exceed $5 \times 10^{13}$ K. They also reported a brightness temperature observed in PKS 1413+135 of $4 \times 10^{12}$ K in 2001. For  these reasons it would not be surprising if the components that are being millilensed in PKS 1413+135 have brightness temperatures in the range of $10^{12}-10^{14}$K. We do not consider microlensing due to a lens of mass $\sim 10 M_{\odot}$ since this would  imply  a brightness temperature of $10^{15}$ K, as can be seen from equation (5), Fig. 6, and Appendix B of Paper 1. Note that (i) if the emission-frame brightness temperature is $10^{12}$K then the variability Doppler factors derived for component D8 given in the two MJD ranges in Table \ref{tab:vardop} drop to $4.0\pm0.3$  for z=0.247 and $4.8\pm0.3$ for z=0.5 in the first MJD window, and $6.2\pm0.4$ for z=0.247 and $7.4\pm0.5$ for z=0.5 in the second MJD window; and (ii) at the higher end of this temperature range the Einstein radius, and hence the typical component separations, would only be
$\sim 20$ microarcseconds.  

Mild chromaticity in SAVs is possible and most likely due to a fast moving shock in a slower underlying flow. In this scenario, we expect the lower frequencies to have a wider U-shape event reaching minimum at a later time, as is observed and quantified in SAV1,SAV2, and SAV4+5. If the millilensing hypothesis is correct, the frequency dependence of SAVs in PKS 1413+135 would provide an unprecedented laboratory to explore jet emission processes. Future work exploring this should include finite source effects since blazar jet emission regions can also vary in size as a function of frequency.  

We have developed a versatile fitting pipeline for millilensing events in light curves. While it is used here for a binary lens model with external convergence and shear to fit SAVs, in practice it can be used for any millilensing light curve event and lens model, so long as the forward lens model magnification can be calculated efficiently. The ability of the module to simultaneously fit millilensing events in multiple frequencies make it ideal to study AGN millilensing with future surveys such as the Legacy Survey of Space and Time (LSST) of the Vera Rubin Observatory \citep[Chapter~10]{LSST2009}. The code can be currently used to model single point lenses, binary point lenses and singular isothermal spheres all with optional external convergence and shear. We are currently extending the code to include finite source size effects. 
 
 The second major component of this study is that the hypothesis that SAVs in PKS 1413+135 are caused by gravitational millilensing has survived the test of two additional SAVs we have identified in the light curves. {Although the joint fitting of SAVs 1,2,4,5 is lacking in some areas, for example the frequency dependence of SAV1 and the SAV2 minimum discussed above, our simple binary lensing model is able to capture the main features of four heterogeneous SAVs.}
 If the millilensing hypothesis is correct, then the PKS 1413+135 plus intervening Seyfert2 galaxy system  is unique to the best of our knowledge.  No other blazar has an intervening edge-on active galaxy in which the blazar core is projected on the sky only $13\pm4$ mas ($52\pm 16$ pc) from the center of activity of the intervening active galaxy. With this unique system we are able  to probe this blazar jet in unprecedented detail on microarcsecond scales, which is otherwise accessible only with a mm wavelength VLBI array in space.
 
We conclude by listing the principal facts that support the gravitational millilensing hypothesis:

\begin{enumerate}
\itemsep0em 
\item Symmetry: The SAV (SAV4), which first drew our attention to this phenomenon, is extraordinarily symmetric, as can be seen in Fig. \ref{plt:symmetry}
\item Repetition: Such symmetric features are rare in blazar light curves (Paper 1) and therefore to find four of them in the light curve of the same blazar by random chance is extremely unlikely, but these repetitions have a natural explanation on the gravitational millilensing hypothesis.
\item Achromaticity: SAVs are near-achromatic from a few GHz to hundreds of GHz.
\item Speed: The speeds of the lensed components are relativistic and in the same range as that covered by the observed speeds of components in PKS 1413+135. There is no {\it a priori\/} reason why this should be the case. In principle, lensed components over a wide  range of speeds could have been detected in the OVRO 15 GHz light curves with the 3-7 day cadence we have maintained over the last 12 years.
\item We have been able to fit  four SAVs seen in PKS 1413+135 over a wide frequency range with the same unvarying lens model.
\item Host: The line of  sight through the intervening spiral galaxy passes within $52\pm 16$ pc of the galactic nucleus and has a path length of tens of kpc through the disc of the edge-on galaxy, so that the cross-section for lensing by mass condensates in the $10^2-10^4 M_\odot$ mass range is unusually high.
\item Potential millilens: \citet{2002AJ....124.2401P} have presented strong evidence of a GMC along the line of sight to the radio core of PKS 1413+135. Thus, there is independent evidence of a mass condensation along the line of sight which {could} have the surface density required for strong millilensing.
\end{enumerate}

If the millilensing hypothesis is correct, then the additional resolution it provides is enabling us to probe the jet with unprecedented resolution, and is probing the three-dimensional jet structure and providing support for the ``fast spine - slow sheath'' model for relativistic jets. We are continuing the high-cadence monitoring of this objects at multiple radio frequencies and hope for another SAV in the next few years that can be followed up with a wide range of observations including millimeter VLBI to search for multiple images.

\acknowledgements
The OVRO 40\,m program was supported by private funding from the California Institute of Technology and the Max Planck Institute for Radio Astronomy, and by NASA grants NNG06GG1G, NNX08AW31G, NNX11A043G, and NNX13AQ89G from 2006 to 2016 and NSF grants AST-0808050, and AST-1109911 from 2008 to 2014.  T. H. was supported by the Academy of Finland projects 317383, 320085, and 322535. S.K. acknowledges support from the European Research Council (ERC) under the European Unions Horizon 2020 research and innovation programme under grant agreement No.~771282. W.M. acknowledges support from ANID projects Basal AFB-170002 and PAI79160080. R.R. acknowledges support from ANID Basal AFB-170002, and ANID-FONDECYT grant 1181620.  
The Submillimeter Array is a joint project between the Smithsonian Astrophysical Observatory and the Academia Sinica Institute of Astronomy and Astrophysics and is funded by the Smithsonian Institution and the Academia Sinica. This research has used data from the University of Michigan Radio Astronomy Observatory, which has been supported by the University of Michigan and by a series of grants from the National Science Foundation, most recently AST-0607523, and from NASA Fermi G. I. grants NNX09AU16G, NNX10AP16G, NNX13AP18G, and NNX11AO13G.
The MOJAVE project was supported by NASA-{\it Fermi} grants 80NSSC19K1579, NNX15AU76G and NNX12A087G.
The Very Long Baseline Array and the National Radio Astronomy Observatory are facilities of the National Science Foundation operated under cooperative agreement by Associated Universities, Inc. 
This work made use of the Swinburne University of Technology software correlator \citep{2011PASP..123..275D}, developed as part of the Australian Major National Research Facilities Programme and operated under licence.
AMI is supported by the Universities of Cambridge and Oxford, and
acknowledges support from the European Research Council under grant
ERC-2012-StG-307215 LODESTONE.

\facilities{Mets\"ahovi, OVRO-40m, SMA, UMRAO, MRAO}

\bibliographystyle{aasjournal}
% Use the LaTeX power, use bibtex properly.
\bibliography{bibliography} %graphy.bib}%,bibliography_export.bib}

\end{document}